\documentclass[a4paper,11pt]{article}
\usepackage{jheppub}
\usepackage{epsfig,fancyvrb,makeidx,latexsym,relsize}

\def\mET{E_T \hspace{-1.2em}/~~}

\def\lsim{\mathrel{\rlap{\lower4pt\hbox{\hskip1pt$\sim$}}
    \raise1pt\hbox{$<$}}}    
\def\gsim{\mathrel{\rlap{\lower4pt\hbox{\hskip1pt$\sim$}}
    \raise1pt\hbox{$>$}}}                
\preprint{RECAPP-HRI-2013-013, MAN/HEP/2013/10}
\title{Invisible Higgs Decay in a Supersymmetric Inverse Seesaw Model with 
Light Sneutrino Dark Matter}
\author[a]{Shankha Banerjee,}
\author[b]{P. S. Bhupal Dev,}
\author[c]{Subhadeep Mondal,}
\author[a]{Biswarup Mukhopadhyaya,}
\author[c]{Sourov Roy}
\affiliation[a]{Regional Centre for Accelerator-based Particle Physics, Harish-Chandra Research Institute, Chhatnag Road, Jhusi, Allahabad 211019, India}
 \affiliation[b]{Consortium for Fundamental Physics, School of Physics and Astronomy,\\ University of Manchester, Manchester, M13 9PL, United Kingdom}
\affiliation[c]{Department of Theoretical Physics,
Indian Association for the Cultivation of Science, 
2A \& 2B Raja S.C. Mullick Road, Kolkata 700032, India}
\emailAdd{shankha@hri.res.in}
\emailAdd{Bhupal.Dev@hep.manchester.ac.uk}
\emailAdd{tpsm2@iacs.res.in}
\emailAdd{biswarup@hri.res.in}
\emailAdd{tpsr@iacs.res.in}
\abstract{
Within the framework of a constrained Minimal Supersymmetric Standard Model 
(cMSSM) augmented by an MSSM singlet-pair sector to account for the non-zero 
neutrino masses by inverse seesaw mechanism, the lightest supersymmetric 
particle (LSP) can be a mixed sneutrino with mass as small as 50 GeV, 
satisfying all existing constraints, thus qualifying as a light dark matter 
candidate. We study the possibility of the lightest neutral Higgs boson in 
this model decaying invisibly into a pair of sneutrino LSPs, thereby giving 
rise 
to novel missing energy signatures at the LHC. We perform a two-parameter 
global 
analysis of the LHC Higgs data available till date to 
determine the optimal invisible Higgs branching fraction in this scenario, and obtain a 
$2\sigma~(1\sigma)$ upper limit of 0.25 (0.15). 
A detailed cut-based analysis 
is carried out thereafter, demonstrating the viability of our proposed signal 
vis-a-vis backgrounds at the LHC.
}
\begin{document}
\maketitle
\section{Introduction}
After the recent discovery~\cite{ATLAS, CMS} of a Higgs boson with mass around 125 GeV, a major goal is to establish whether 
it is `the' Standard Model (SM) Higgs boson or a first glimpse of some Beyond Standard Model (BSM)  physics 
at the LHC. A precise determination of the discovered Higgs boson 
characteristics will be crucial in 
resolving some of the outstanding issues of the SM, and in particular, 
understanding the mechanism of electroweak symmetry breaking and its 
relationship to the BSM. 
The experimental results so far ~\cite{ATLAS1, ATLAS2, CMS1, ATLAS3} show 
no significant deviation from the SM Higgs sector expectations, 
and already put severe constraints on various new physics models 
(see, for instance,~\cite{Corbett:2012dm, Banerjee:2012xc, Belanger:2012gc, Cheung:2013kla, Falkowski:2013dza, Ellis:2013lra, Djouadi:2013qya, Giardino:2013bma}). However, they still do not 
exclude the possibility of a non-standard Higgs boson.     
 
A precise measurement of the total decay width of the Higgs boson ($h$) 
through its line shape is very difficult at the LHC due to its tiny value: 
for the SM with $m_h=125$ GeV, $\Gamma_h= 4.07$ MeV~\cite{Dittmaier:2011ti}. Hence, a better way to identify a non-standard Higgs boson is by studying its 
non-standard decay modes (for a review, see e.g.,~\cite{Chang:2008cw}). 
This is also crucial in case of a statistically 
significant discrepancy between the measured and SM expected Higgs signal strengths which 
could be due to either suppression or enhancement of the Higgs production cross section as well as its partial decay widths.     

A particularly interesting non-standard Higgs decay which is very sensitive to large 
BSM contributions is its invisible decay mode~\cite{Shrock:1982kd}, 
since the SM invisible Higgs branching ratio (BR) is very small: BR$(h\to ZZ^*\to 4\nu)\simeq 0.001$~\cite{Denner:2011mq}. 
Dedicated searches for the Higgs decay into invisible final states were performed at the LEP~\cite{Abdallah:2003ry}, and no signal was found for Higgs mass 
up to 114.4 GeV. 
The LHC prospects of determining the invisible Higgs BR have been analyzed 
in Refs.~\cite{Gunion:1993jf, Frederiksen:1994me, 
Eboli:2000ze, Kersevan:2002zj, Belotsky:2002ym,  Godbole:2003it, Davoudiasl:2004aj,Zhu:2005hv, 
Bai:2011wz, Ghosh:2012ep}. The current experimental limits for a 125 GeV 
invisible Higgs BR are < 0.65 from ATLAS~\cite{ATLAS-011} and < 0.75 from CMS~\cite{cmsinv} at 95\% CL derived from the 
direct search $pp\to Zh\to \ell \ell \mET$. Global fits to the existing LHC data provide a stronger constraint 
on BR$_{\rm inv}<0.28$ at 95\% CL~\cite{Giardino:2013bma} 
(for other recent global fits, see~\cite{Espinosa:2012vu, Belanger:2013kya}). 
 %
 
From a phenomenological point of view, the compelling evidence for the 
existence of dark matter (DM) and its `WIMP-miracle explanation' (for a review, see e.g.,~\cite{Bertone:2004pz}) suggest that 
given suitable mass and unsuppressed coupling to the Higgs, the invisible decay to DM could be significant. 
In fact, this can occur in many well-motivated BSM scenarios, e.g., MSSM with neutralino 
DM~\cite{Griest:1987qv,Ananthanarayan:2013fga,Djouadi:1992pu, Belanger:2001am, AlbornozVasquez:2011aa, Arbey:2011aa, Desai:2012qy, Dreiner:2012ex}, 
models with extended scalar 
sector~\cite{Binoth:1996au, Burgess:2000yq, Andreas:2010dz, Raidal:2011xk, Pospelov:2011yp, Mambrini:2011ik}, 
Majoron models~\cite{Romao:1992zx, Joshipura:1992hp, Choudhury:1993hv, Hirsch:2004rw, Ghosh:2011qc}, 
large extra dimension~\cite{Datta:2004jg, Dominici:2009pq}, etc. 
The possibility of Higgs decaying to DM has gained renewed interest 
in view of the recent claims from some DM direct detection experiments such as 
DAMA/LIBRA~\cite{dama}, CoGeNT~\cite{cogent}, CRESST-II~\cite{cresst}, and more recently CDMS-II~\cite{cdms}, 
in favor of a light DM in the mass range 
$5$ - $50~{\rm GeV}$, with a large DM-nucleon scattering cross section of $10^{-5}-10^{-7}$ pb. This provides 
a strong motivation to examine the invisible Higgs decays in some BSM scenarios accommodating a light DM. 

Due to various well-known theoretical reasons (see e.g.,~\cite{baer}), 
low-scale supersymmetry (SUSY) remains as one of the most attractive BSM scenarios, in spite of the null results from SUSY searches at the LHC so far~\cite{atlas-susy, cms-susy}. In $R$-parity conserving SUSY models, the lightest supersymmetric particle (LSP), if electrically neutral,  provides a natural WIMP DM candidate (for a review, see~\cite{susy-dm}). In the Minimal Supersymmetric 
Standard Model (MSSM), the lightest neutralino is the usual DM candidate, as the other viable candidate, namely, the scalar superpartner of the left-handed (LH) neutrino (the LH sneutrino), is strongly disfavored due to constraints from DM relic density and direct detection as well as the invisible decay width of the $Z$-boson~\cite{Falk:1994es, Hebbeker:1999pi}. However, in the minimal Supergravity (mSUGRA)/constrained MSSM (cMSSM)~\cite{msugra} with gaugino and sfermion mass unification, 
the recent LHC data disfavor a light neutralino mass below about 200 GeV, as demonstrated by the global fits~\cite{Buchmueller:2012hv, Strege:2012bt, Cabrera:2012vu, Kowalska:2013hha}, thus excluding the possibility of Higgs decaying to neutralino DM.\footnote{In a more general version of the MSSM it is still possible to have a light neutralino satisfying all the experimental constraints~\cite{Choudhury:1999tn, Dreiner:2009ic} though these cases turn out to be highly fine-tuned 
(see e.g.,~\cite{Grothaus:2012js, Boehm:2013qva}).} For other implications of the recent experimental results for mSUGRA, see e.g.,~\cite{Nath:2012fa}.

On the other hand, the neutrino oscillation data require at least two of the three SM neutrinos to have a tiny but non-zero mass (for a review, see e.g.,~\cite{GonzalezGarcia:2007ib}) which calls for some new physics beyond SM/MSSM. Thus, 
it would be interesting if a simple extension of the MSSM to explain the 
neutrino oscillation data can also accommodate a light DM 
candidate while satisfying all the existing experimental constraints. Such a 
scenario was recently studied in Ref.~\cite{BhupalDev:2012ru} within the framework of cMSSM supplemented by a SM singlet-pair sector to explain the non-zero 
neutrino masses and mixing by a low-scale inverse seesaw mechanism~\cite{inverse1, inverse2}. It was shown that in contrast with the pure cMSSM scenario, this allows a light DM in the form of a mixed sneutrino with mass around $m_h/2$, required to have a large annihilation rate via $s$-channel Higgs resonance. 

Due to the large Yukawa couplings allowed in the 
model which are responsible for an efficient annihilation of the sneutrino DM, 
the lightest $CP$-even Higgs boson can have a large invisible 
branching ratio to sneutrino final states. This in turn leads to novel 
missing energy signatures at the LHC. Here we analyze this possibility in detail by performing a two-parameter global fit with the latest LHC Higgs data to determine the 
optimal invisible Higgs branching ratio allowed in this model, and find a $2\sigma~(1\sigma)$ upper limit of 0.25 (0.15). This 
in turn puts an upper limit of ${\cal O}(0.1)$ on the Dirac Yukawa coupling in the model. We further show that the model parameter 
space allowed by the invisible Higgs decay constraints can be completely ruled out in case of null results at the next generation DM direct detection 
experiments 
such as LUX and XENON1T. Finally, we select a few benchmark 
points satisfying all the experimental constraints, and carry out a detailed cut-based analysis, 
demonstrating the viability of our proposed signal in two Higgs production channels, namely, vector boson fusion (VBF) and associated production with $Z$, vis-a-vis SM backgrounds at $\sqrt s$=14 TeV LHC. We find 
that a signal significance of $3\sigma$ can be achieved in the VBF channel with an integrated luminosity as low as $200~{\rm fb}^{-1}$, whereas in the $Zh$ channel it requires a luminosity of at least $600~{\rm fb}^{-1}$ for our chosen benchmark points.

The paper is organized in the following way. In Section 2 we
give a brief description of the model. 
In Section 3, we scan the model parameter space to select a few benchmark points 
for a viable light sneutrino DM candidate. Then we perform a global $\chi^2$-analysis with
the available LHC Higgs data to obtain the $1\sigma$ and $2\sigma$ allowed ranges of invisible Higgs branching fraction. In Section 4 we present a collider analysis for the invisible Higgs decay signature 
at the LHC, focusing on two of its production channels, namely, VBF and $Zh$, for a few chosen 
benchmark points satisfying all the experimental constraints. Our conclusions are given in Section 5. In the Appendix, we list all the ATLAS and CMS Higgs data sets used in our global analysis.
\section{An Overview of the Model}
In the supersymmetric version of the inverse seesaw mechanism~\cite{inverse1, inverse2}, 
all the light neutrino masses can be generated at tree-level by adding three pairs of SM singlet 
superfields: $\hat{N}^c_i$ and $\hat{S}_i$ (with $i=1,2,3$) having lepton number $-1$ and $+1$ respectively. 
Thus the sneutrino LSP is in general a linear combination of the superpartners of the LH neutrino and the 
singlet fermions. Several embeddings of this set up have been discussed in the literature within the MSSM 
gauge group~\cite{Arina:2008bb} as well as with extended gauge symmetries such as 
$SU(2)_L\times SU(2)_R\times U(1)_{B-L}$~\cite{Dev:2009aw, Dev:2010he, An:2011uq}, 
$SU(2)_L\times U(1)_Y\times U(1)_{B-L}$~\cite{Khalil:2011tb, Basso:2012ti} and 
$SU(2)_L\times U(1)_Y\times U(1)_R$~\cite{DeRomeri:2012qd}. In this paper, we choose to work within 
the MSSM gauge group and take a hybrid approach for the model 
parameters similar to that in Refs.~\cite{Arina:2008bb,BhupalDev:2012ru} to find suitable benchmark points, 
i.e. a low energy input for the MSSM singlet fermion sector and for the lepton-number violating soft 
SUSY-breaking sector while a top-down approach for the MSSM sparticle spectrum with mSUGRA boundary 
conditions at the high scale without necessarily imposing any features of a specific Grand Unified 
Theory (GUT) framework. The mSUGRA boundary conditions for the MSSM sector enables us a direct 
comparison with the pure cMSSM case for its collider phenomenology~\cite{BhupalDev:2012ru}. The low-energy inputs 
for the singlet sector are chosen to satisfy all the low-energy constraints 
in the lepton sector. It is reasonable to choose them directly at 
the SUSY-breaking scale since the Renormalization Group (RG) running effects from the singlet sector on the pure cMSSM sector are expected to be small (as can be seen, for instance, from the RG equations in Ref.~\cite{Dev:2010he} in the context of a particular $SO(10)$ GUT model), and hence, it is equivalent to choosing a corresponding set of RG-evolved high-energy inputs, thus making our analysis independent of any specific GUT embedding. Henceforth, we will refer to this hybrid model generically as the 
Supersymmetric Inverse Seesaw Model (SISM).   

The SISM superpotential is given by
\begin{eqnarray}
	{\cal W}_{\rm SISM}={\cal W}_{\rm MSSM}+\epsilon_{ab}y_\nu^{ij}\hat{L}^a_i\hat{H}^b_u \hat{N}^c_j+ M_{R_{ij}}
\hat{N}^c_i\hat{S}_j+\mu_{S_{ij}}\hat{S}_i\hat{S}_j \, ,
\label{sup}
\end{eqnarray}
$\mu_S$ being the only (tiny) source of lepton number violation in the superpotential. The soft SUSY-breaking Lagrangian is given by  
\begin{eqnarray}
	{\cal L}_{\rm SISM}^{\rm soft} &=& {\cal L}_{\rm MSSM}^{\rm soft} -\left[m_N^2\widetilde{N}^{c^{\dag}}\widetilde{N}^c+
	m_S^2\widetilde{S}^\dag\widetilde{S}\right]\nonumber\\
	&& -\left[\epsilon_{ab}A_\nu^{ij}\widetilde{L}^a_i\widetilde{N}^c_jH_u^b+B^{ij}_{M_R}\widetilde{N}^c_i\widetilde{S}_j+
	B_{\mu_S}^{ij}\widetilde{S}_i\widetilde{S}_j+{\rm h.c.}\right]. 
\label{soft}
\end{eqnarray}
As a result of the LH neutrinos mixing with the singlet ones, the tree level 
 neutrino mass matrix is $9\times 9$ in the basis $\{\nu_L, N^c, S\}$: 
\begin{eqnarray}
	{\cal M}_\nu = \left(\begin{array}{ccc}
		{\bf 0} & M_D & {\bf 0}\\
		M_D^T & {\bf 0} & M_R \\
		{\bf 0} & M_R^T & \mu_S
	\end{array}\right),
	\label{eq:mbig}
\end{eqnarray}
where $M_D = v_u y_{\nu}$ is the Dirac neutrino mass matrix, $v_u=v\sin\beta$ being the vacuum expectation value (vev) of the $\hat H_u$ superfield in MSSM, with $v\simeq 174$ GeV.
In the limit $\|\mu_S\|\ll \|M_R\|$ (where $\|M\|\equiv \sqrt{{\rm Tr}(M^\dag M)}$), we can extract the $3\times 3$ 
light neutrino mass matrix as 
\begin{eqnarray}
	M_\nu = \left[M_DM_R^{T^{-1}}\right]\mu_S\left[(M_R^{-1})M_D^T\right]+{\cal O}(\mu_S^2) \equiv F\mu_S F^T+{\cal O}(\mu_S^2) \, .
	\label{eq:vmass}
\end{eqnarray}
As can be seen from Eq.~(\ref{eq:vmass}), the smallness of neutrino mass
now additionally depends on the small lepton-number violating parameter 
$\mu_S$ instead of just the smallness of the Dirac mass $M_D$ and/or heaviness of $M_R$ as 
in the canonical type-I seesaw case~\cite{seesawa, seesawb, seesawc, seesawd, seesawe}. For 
$\mu_S\sim {\cal O}({\rm keV})$, we can easily bring down $M_R$ to ${\cal O}$(TeV) 
range even with comparatively large Dirac Yukawa couplings of ${\cal O}(0.1)$, 
thus leading to a rich collider phenomenology~\cite{BhupalDev:2012ru, Chen:2011hc, Mondal:2012jv, Das:2012ze, Bandyopadhyay:2012px} as well as observable lepton flavor violation (LFV)  
effects~\cite{Dev:2009aw, Deppisch:2004fa, Deppisch:2005zm, Malinsky:2009gw, Malinsky:2009df, Hirsch:2009ra, Abdallah:2011ew, Abada:2011hm, Hirsch:2012ax, 
Abada:2012cq, Awasthi:2011aa, Awasthi:2013ff}. 

\subsection{Fitting Neutrino Oscillation data}
The effective light neutrino mass matrix is usually 
diagonalized by the unitary Pontecorvo-Maki-Nakagawa-Sakata (PMNS)
matrix. But due to its mixing with heavy neutrinos in the matrix structure of 
${\cal M}_\nu$ in Eq.~(\ref{eq:vmass}), the light neutrino mixing matrix will 
receive additional non-unitary contributions. Thus, the full (non-unitary) 
light neutrino mixing matrix ${\cal U}$ diagonalizing the light neutrino mass matrix in Eq.~(\ref{eq:vmass}) has to be derived from the $9\times 9$ 
unitary matrix ${\cal V}$ diagonalizing the full mass matrix given in
Eq.~(\ref{eq:mbig}), i.e,
\begin{eqnarray}
{\cal V}{\cal M}_\nu {\cal V}^T = {\rm diag}(m_i,m_{R_j}),~ ~(i=1,2,3;~j=1,2,...,6), 
\end{eqnarray}
and by decomposing it into the blocks
\begin{eqnarray}
{\cal V}_{9\times 9}=\left(\begin{array}{cc}
{\cal U}_{3\times 3} & {\cal K}_{3\times 6}\\
{\cal K}'_{6\times 3} & {\cal N}_{6\times 6}
\end{array}\right).
\label{eq:diagfull}
\end{eqnarray}
For $\|M_D\|\ll \|M_R\|$, it is sufficient to expand ${\cal U}$ up to leading 
order in $F=M_D M_R^{T^{-1}}$:
\begin{eqnarray}
{\cal U}\simeq \left({\bf 1} - \frac{1}{2}FF^\dag\right)U \equiv 
({\bf 1}-\eta)U
\label{eq:nonunitary}
\end{eqnarray}
where U denotes the unitary Pontecorvo-Maki-Nakagawa-Sakata (PMNS) matrix that diagonalizes the light neutrino mass matrix
and $\eta=\frac{1}{2}FF^\dag$ is a measure of the non-unitarity.

In order to satisfy the LFV constraints simultaneously with the sneutrino DM relic density constraint, we choose to work with diagonal $M_R$ and $M_D$, and  
accordingly fit $\mu_S$ to be consistent with the neutrino oscillation data. We use the following global fit 
values for the oscillation parameters~\cite{valleglobal}: 
\begin{eqnarray}
  && \Delta m^2_{21} = (7.62\pm 0.19)\times 10^{-5}~{\rm eV}^2,~ ~ \Delta m^2_{31} = (2.53\pm 0.09)\times 10^{-3}~{\rm eV}^2,\nonumber\\
  && \sin^2\theta_{12} = 0.320\pm 0.016,~ ~ \sin^2\theta_{23} = 0.490\pm 0.065,~ ~ \sin^2\theta_{13} = 0.026\pm 0.004.
\label{eq:oscil}
\end{eqnarray} 
\subsection{Sneutrino mass matrix}
In the scalar sector, due to mixing between doublet and singlet
sneutrinos we have an analogous $9\times 9$ complex (or $18\times 18$ real) 
sneutrino mass squared matrix. 
Assuming $CP$ conservation in the soft SUSY-breaking Lagrangian (\ref{soft}),\footnote{The addition of extra $CP$ phases do 
not affect any of the collider aspects studied in this paper; hence they were taken to be zero for simplicity.} 
we can decompose this mass matrix into two 
$9\times 9$ real block-diagonal matrices corresponding to $CP$-even and $CP$-odd
sneutrino states. The corresponding mass term in the Lagrangian looks like
\begin{eqnarray}
 {\cal L}_{\tilde\nu} = \frac{1}{2} (\phi^R, \phi^I) \left(\begin{array}{cc}
		{\cal M}_+^2 & {\bf 0} \\
		{\bf 0} & {\cal M}_-^2
	\end{array}\right) \left(\begin{array}{c}{\phi^R}\\ {\phi^I}\end{array}\right),
\end{eqnarray}
where $\phi^{R,I} = (\widetilde\nu^{R,I}_{L_i},\widetilde N^{c^{R,I}}_j,\widetilde S^{R,I}_k)~ 
(i,j,k = 1, 2, 3)$ and  
\begin{eqnarray}
{\cal M}_\pm^2 = \left(\begin{array}{ccc}
		m_{\tilde L}^2+M_DM_D^T+\frac{1}{2}m_Z^2\cos 2\beta & \pm(v_uA_\nu-\mu M_D\cot\beta) & M_DM_R\\
		\pm(v_uA_\nu-\mu M_D\cot\beta)^T & m_N^2+M_RM_R^T+M_D^TM_D & B_{M_R}\pm M_R\mu_S\\
		M_R^TM_D^T & B^T_{M_R}\pm \mu_SM_R^T & m_S^2+\mu_S^2+M_R^TM_R\pm B_{\mu_S}
	\end{array}\right), \nonumber
	\label{eq:svmass}
\end{eqnarray}
where $m_{\tilde L}^2$ denote the soft SUSY-breaking mass squared term for $SU(2)_L$-doublet sleptons. The real 
symmetric $CP$-even and $CP$-odd mass squared 
matrices ${\cal M}_\pm^2$ can be diagonalized by $9\times 9$ orthogonal 
matrices ${\cal G}_\pm$ as follows:
\begin{eqnarray}
 {\cal G}_\pm {\cal M}^2_\pm {\cal G}^T_\pm = {\rm diag} \left(m^2_{\widetilde\nu^{R,I}_i}\right)~~~~(i=1, 2,\cdots, 9).
\end{eqnarray}
The corresponding eigenvalues of ${\cal M}^2_\pm$ are
almost degenerate in nature, with the degeneracy between $\widetilde\nu^{R}_i$ 
and $\widetilde\nu^{I}_i$ lifted only due to the small lepton number breaking parameter $\mu_S$. We will choose some 
benchmark points for which the lightest 
sneutrino mass eigenstate is the LSP, and will serve as a light DM candidate. 
\section{Invisible Higgs Decay}
Our goal in this section is to find the prospects of the lightest $CP$-even Higgs boson 
decaying into two light DM particles in the form of sneutrino LSP, 
thereby leading to a missing energy signal at the LHC. In the SISM being discussed here, 
we have 5 mSUGRA parameters $m_0, m_{1/2}, \tan\beta, A_0, {\rm sign} (\mu)$ at high scale and the additional 
inverse seesaw parameters $M_D, M_R, \mu_S, B_{\mu_S}$ and $B_{M_R}$ whose 
input values are chosen at the low scale. For simplicity, we have assumed these low-energy 
neutrino sector parameters to be diagonal (apart from $\mu_s$ whose structure is fixed by neutrino oscillation data) 
so that we can easily satisfy the LFV 
constraints. Also, the trilinear $A_\nu$ term in the soft 
SUSY-breaking Lagrangian which controls the Higgs BR to sneutrinos 
is taken to be $(A_\nu)_{ij}=A_0(y_\nu)_{ij}$. Note that we require a large $A_0$ in order 
to have a large radiative correction to the lightest $CP$-even Higgs mass as required by 
the LHC observation, whereas the Dirac Yukawa coupling $y_\nu$ is also required to be large 
in order to provide an efficient annihilation channel for the sneutrino LSP. These two 
seemingly uncorrelated effects inevitably lead to a large invisible BR for the Higgs in the 
SISM.  
\subsection{Light Sneutrino DM}
\begin{figure}[h!]
\centering
\includegraphics[height=5cm,width=7cm]{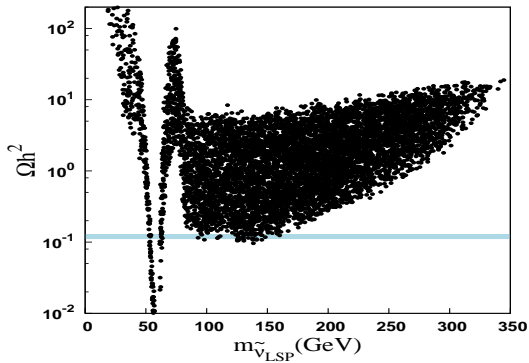}
\caption{Sneutrino relic density as a function of the sneutrino LSP mass for our SISM input parameter scan. 
The horizontal shaded band shows the Planck $3\sigma$ preferred range. }
\label{fig:relic_den}
\end{figure}

It was shown in~\cite{BhupalDev:2012ru} that the observed DM relic density for light 
sneutrino LSPs in the SISM is obtained by resonant enhancement of the annihilation cross section in the 
Higgs-mediated $s$-channel process: $\widetilde{\nu}_{\rm LSP}\widetilde{\nu}_{\rm LSP}\to f\bar{f}$ 
(where $f$ denotes the SM fermion). This is illustrated in Figure~\ref{fig:relic_den} which was obtained by 
choosing the input parameters in a sample range 
\begin{eqnarray}
&&m_0\in [0.1, 2.5]~{\rm TeV},~m_{1/2}\in [0.65, 2.5]~{\rm TeV},~
A_0 \in [-3, 3]~{\rm TeV},\nonumber\\
&& {\rm diag}(y_\nu)\in [0.01,0.2],~(M_R)_{11}\in [100, 800]~{\rm GeV},
\label{range}
\end{eqnarray}
and for a fixed $\tan\beta=10,~{\rm sign}(\mu)=+1,~(M_R)_{22,33}=1~{\rm TeV},~B_{\mu_S} = 10^{-4}~{\rm
  GeV}^2$, and $B_{M_R}=10^6~{\rm GeV}^2$. We have chosen the mSUGRA parameter ranges shown here keeping in mind the LHC 
  exclusion limits on the cMSSM parameter space~\cite{atlas-susy, cms-susy}. The parameter scan was performed 
  using {\tt SSP}~\cite{Staub:2011dp}, with the SISM 
implemented in {\tt SARAH}~\cite{sarah}, and the sparticle spectrum was generated 
using {\tt SPheno}~\cite{spheno}, while DM relic density was calculated using {\tt micrOMEGAs}~\cite{micromegas}. 
All the points shown in Figure~\ref{fig:relic_den} are required to have the lightest $CP$-even Higgs mass in the range 
$125\pm 2$ GeV to be consistent with the latest LHC Higgs data~\cite{ATLAS1, CMS1}. The horizontal blue band indicates 
the $3\sigma$ preferred range from Planck data: 
$\Omega h^2=0.1199\pm 0.0081$~\cite{Ade:2013zuv}. It is clear that for the sneutrino 
LSP mass below $W$-boson mass, the observed DM relic density is obtained only near the 
Higgs-resonance region, thus requiring the sneutrino DM mass in the SISM to be around $m_h/2$. The other possible resonance around $m_Z/2$ is suppressed in this case due to small mixing between the $SU(2)_L$-doublet and singlet neutrinos, as required to satisfy the $Z$-invisible decay width constraint from LEP~\cite{LEP}. Note that in Fig.~\ref{fig:relic_den}, the observed 
relic density can also be satisfied for sneutrino LSP in the 80 - 200 GeV mass range due to its large annihilation rate into $WW,ZZ$ and $hh$ final states. Since our main focus in this paper is on light sneutrino DM and Higgs invisible decay, we do not consider this mass range in our subsequent analysis.    

\begin{figure}[htb]
\centering
\includegraphics[height=5cm,width=7cm]{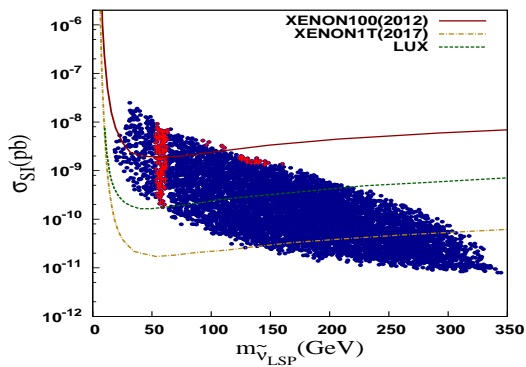}
\caption{Spin-independent direct detection cross section as a function of the sneutrino LSP 
mass for our SISM parameter scan. The red (+) points satisfy relic density $\lsim$ 0.13. 
The current experimental limit from XENON100 and the projected limits from LUX and XENON1T are also shown. }
\label{fig:sigma_si}
\end{figure}

The same interaction that leads to the Higgs-mediated $s$-channel annihilation of the sneutrino DM in our model also leads to a direct detection signal via $t$-channel Higgs exchange. 
In Figure~\ref{fig:sigma_si} we have plotted the spin-independent DM-nucleon scattering 
cross section predictions as a function of the sneutrino LSP mass for the corresponding points in Figure~\ref{fig:relic_den}. We also show the subset of points satisfying the relic density constraints. The solid line indicates the current limit 
from XENON100 data~\cite{Aprile:2012nq}. We also show the projected limits from XENON1T~\cite{Aprile:2012zx} and LUX~\cite{Akerib:2012ys} experiments. As evident from the plot, a few 
of the allowed points are already ruled out by the XENON100 data, while {\it all} of the 
low-mass points satisfying the relic density constraints can be ruled out by LUX and XENON1T projected limits in case of a null result.  

From Figures~\ref{fig:relic_den} and \ref{fig:sigma_si} we infer that it is indeed possible 
to have the lightest Higgs boson decaying into two sneutrino LSPs, while 
satisfying the DM relic density and direct detection constraints. We have also checked that 
all the points shown in Figures~\ref{fig:relic_den} and \ref{fig:sigma_si} are well below 
the current indirect detection cross section limits from Fermi-LAT~\cite{Ackermann:2011wa, Ackermann:2012qk}.   

\subsection{The invisible decay width and current data}
In order to ascertain how much invisible BR of the Higgs is allowed in our model, we perform a global analysis with all the LHC Higgs data available so far (see Appendix). Since 
the neutrino sector parameters of the SISM do not affect the Higgs production or decay rates into the SM final states, and only affect its invisible decay into sneutrino final states, we can parametrize their effect in terms of a single free parameter, namely, 
the invisible BR $\varepsilon$ which relates the visible and invisible partial widths of the Higgs boson as
\begin{equation}
\Gamma_{\rm inv}  =  \frac{\varepsilon}{1-\varepsilon} \sum\Gamma_{\rm vis} \, .
\end{equation}
For the MSSM sector of the SISM, we choose a few benchmark points (BPs) by fixing $m_0, m_{1/2}$ and  $A_0$ as shown in Table~\ref{tab:mssm_scan}, and vary 
the remaining parameter, namely, $\tan\beta$ to compute the low-energy SUSY spectrum using 
{\tt SPheno}~\cite{spheno}. The benchmark points given in 
Table~\ref{tab:mssm_scan} were selected from the sample scan ranged over the values given in Eq.~(\ref{range}) by requiring them to 
satisfy the constraints coming from higgs and squark-gluino mass bounds.
We have fixed the sign of the MSSM $\mu$-parameter to be $+1$ throughout our analysis since $\mu<0$ is strongly disfavored by the 
muon anomalous magnetic moment as well as by the $B\to X_s\gamma$ branching ratio. Note that all the benchmark points shown in 
Table~\ref{tab:mssm_scan} require an electroweak fine-tuning at the percent level, which is mandatory given the current LHC data 
(see e.g.,~\cite{Baer:2012mv}). For the trilinear term $A_0$, a large negative value is required to obtain the correct Higgs mass 
($m_h=125\pm 2$ GeV) for our choices of $m_0$ and $m_{1/2}$ (which are consistent with the general results from other mSUGRA 
parameter scans, e.g.~\cite{Baer:2012mv}). We have checked that all our benchmark points lead to a stable electroweak vacuum 
and do not lead to charge- and/or color-breaking minima. 
  
\begin{table}[h!]
\begin{center}
\begin{tabular}{||c|c|c|c||}\hline\hline
Input parameter & BP1 & BP2 & BP3 \\ 
\hline\hline
$m_0$ (GeV) & 996.45 & 745.48 & 614.00\\ 
$m_{1/2}$ (GeV)  & 750.00& 1014.17 & 1083.00 \\
$A_0$ (GeV) & $-2858.00$ & $-2775.09$ & $-2600.00$\\
\hline\hline 
\end{tabular}
\end{center}
\caption{The mSUGRA input parameters for three chosen benchmark points.}
\label{tab:mssm_scan}
\end{table}

For each combination of the high-scale parameters given in Table~\ref{tab:mssm_scan}, we perform a 
global analysis in the $\varepsilon$-$\tan\beta$ plane using 10 data points in various Higgs decay channels from the 
published results of 
CMS and ATLAS, as listed in 
Appendix. For each of the variables, with the other one marginalized, we compute the $\chi^2$ 
function, defined as
\begin{equation}
\chi^2 = \sum_{i}\frac{(\mu_i - \hat{\mu_i})^{2}}{(\delta\hat{\mu}_i)^2},
\end{equation}
where $\mu_i$'s are the Higgs signal strengths calculated from the model and
are functions of the model parameters:
\begin{equation}
\mu_i = R_{i}^{\rm prod}\times \frac{R_{i}^{\rm decay}}{R^{\rm width}}\, .
\end{equation}
Here $R_i$'s are the ratios of the model predictions for the Higgs production cross sections and 
partial decay rates for various channels, and similarly $R$ is the ratio of the total width, with the corresponding SM expectations:
\begin{eqnarray}
R_i^{\rm prod} = \frac{\left(\sigma_i^{\rm prod}\right)_{\rm SISM}}{\left(\sigma_i^{\rm prod}\right)_{\rm SM}}\, ,~
R_i^{\rm decay} = \frac{\left(\Gamma_i^{\rm decay}\right)_{\rm SISM}}{\left(\Gamma_i^{\rm decay}\right)
_{\rm SM}}\, ,~
R^{\rm width} = \frac{\left(\Gamma^{\rm width}\right)_{\rm SISM}}{\left(\Gamma^{\rm width}\right)_{\rm SM}}\, , 
\end{eqnarray}
and $\hat{\mu_i}$'s are the experimental best fit values of  the signal strengths as listed in Appendix, $\delta \hat\mu_i$'s 
being their reported $1\sigma$ uncertainty. When the reported uncertainties are 
asymmetric in nature, we consider the  positive uncertainty for $(\mu_i -  \hat{\mu_i}) > 0$
and the negative one for $(\mu_i - \hat{\mu_i}) < 0$. 

\begin{figure}[h!]
\includegraphics[height=5cm,width=7cm]{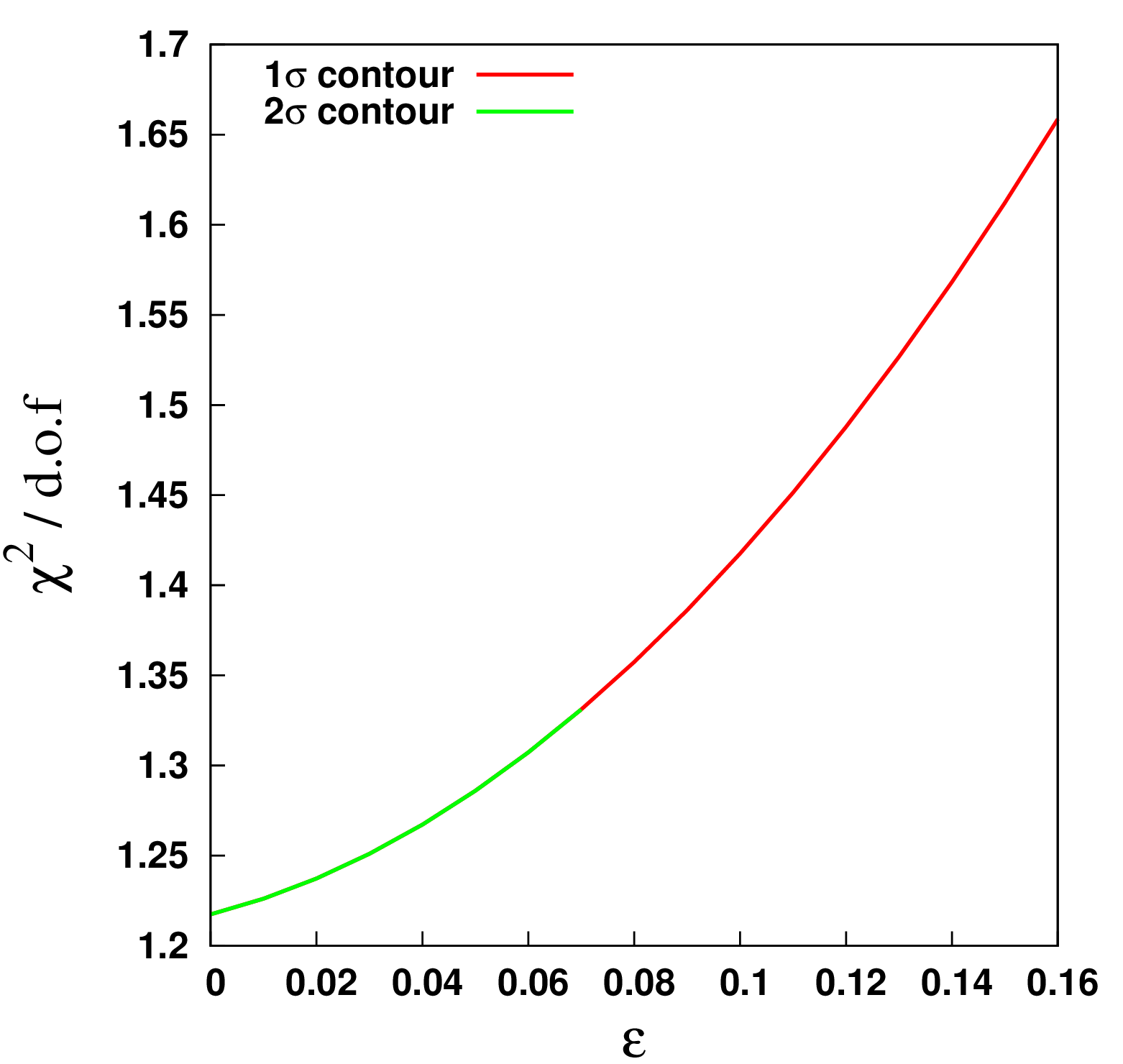} ~~ \includegraphics[height=5cm,width=7cm]{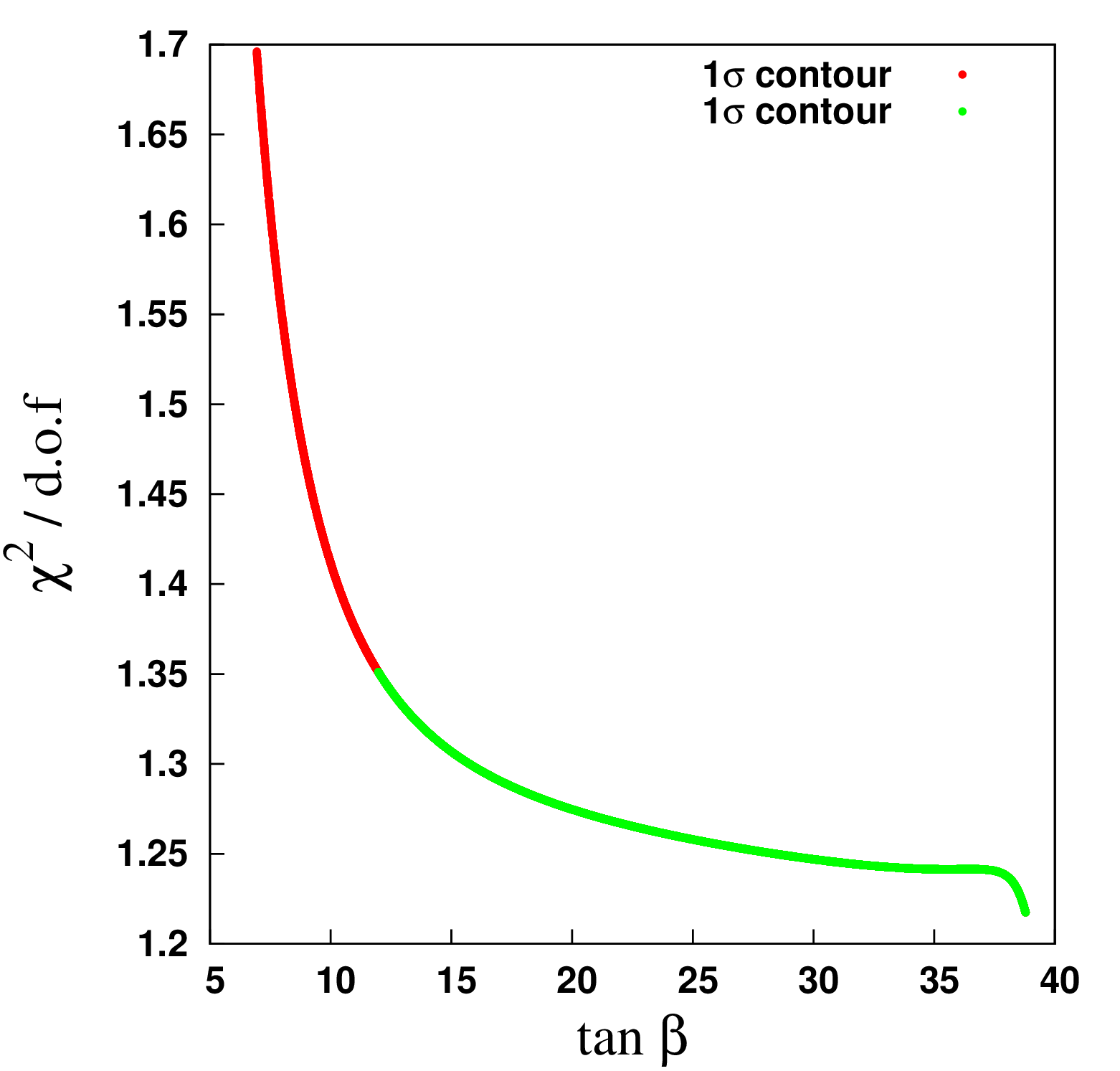}
\caption{$1\sigma$ and $2\sigma$ contours for $\varepsilon$ and $\tan\beta$ from $\chi^2$ minimization obtained for BP1. 
The green line indicates $1\sigma$ reach and the red line indicates $2\sigma$ reach of the parameters.}
\label{fig:BP1_eps_tanb}
\end{figure}

\begin{figure}[h!]
\includegraphics[height=5cm,width=7cm]{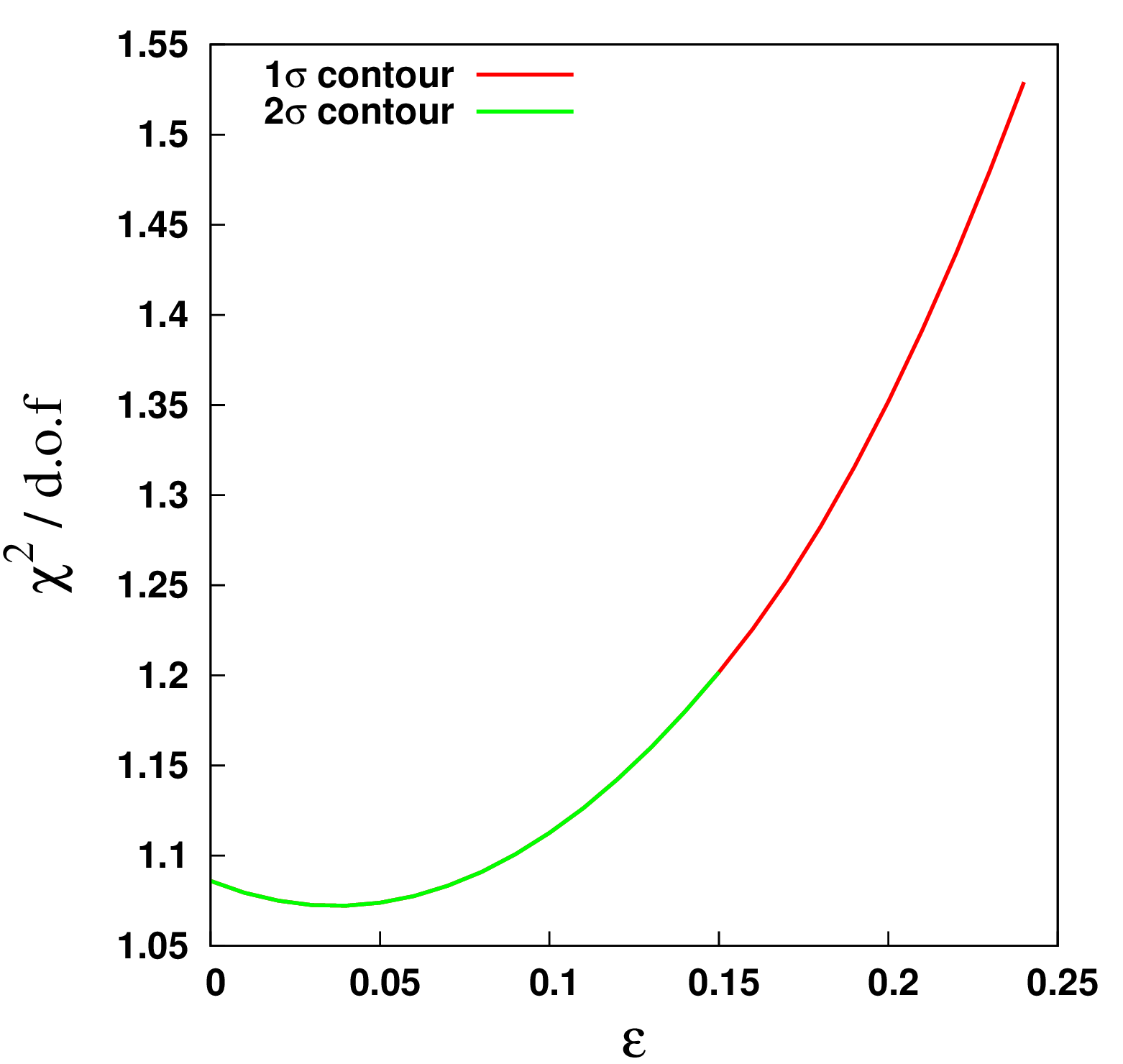} ~~ \includegraphics[height=5cm,width=7cm]{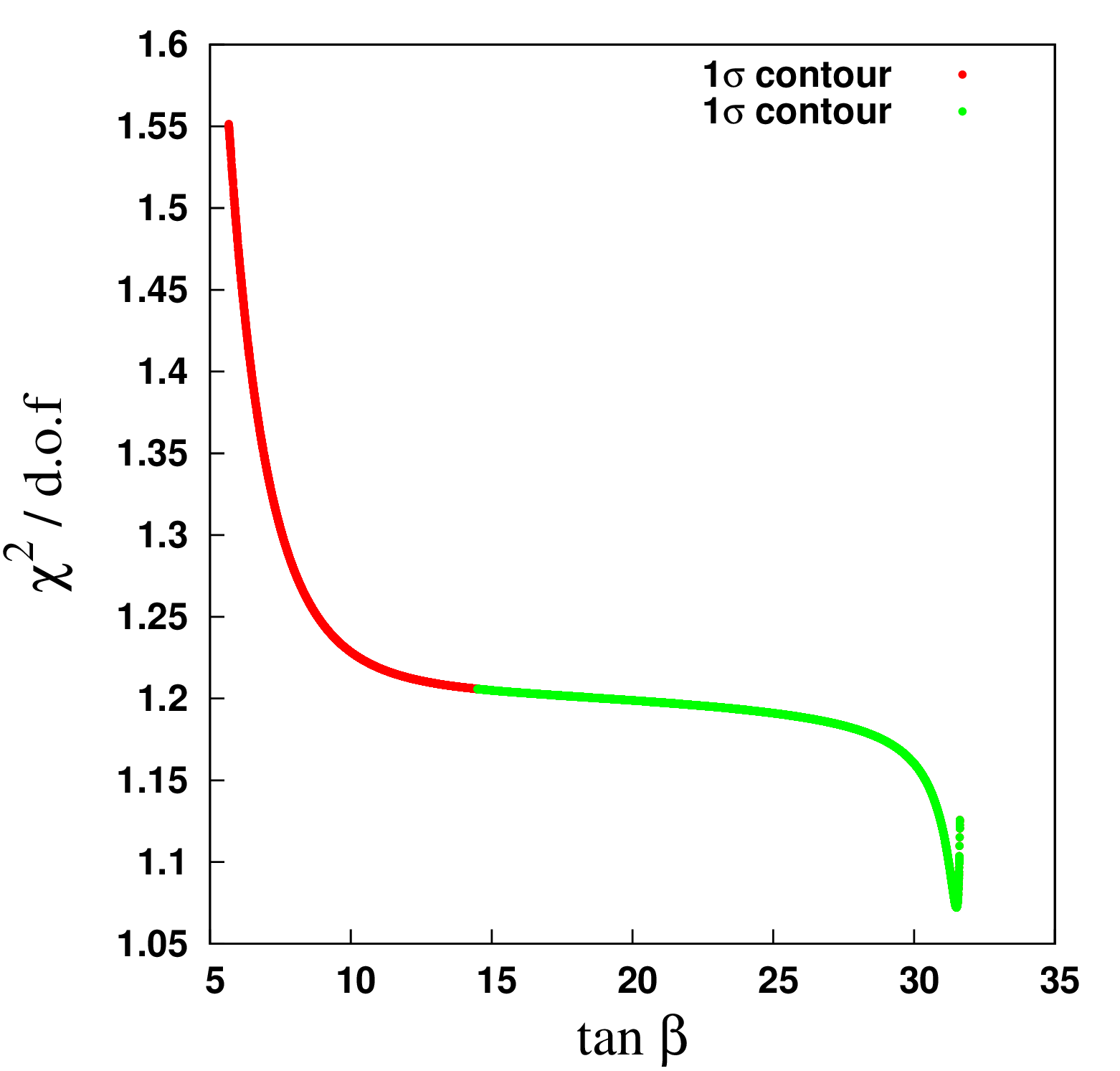}
\caption{$1\sigma$ and $2\sigma$ contours for $\varepsilon$ and $\tan\beta$ from $\chi^2$ minimization obtained for BP2. 
The green line indicates $1\sigma$ reach and the red line indicates $2\sigma$ reach of the parameters.}
\label{fig:BP2_eps_tanb}
\end{figure}

\begin{figure}[h!]
\includegraphics[height=5cm,width=7cm]{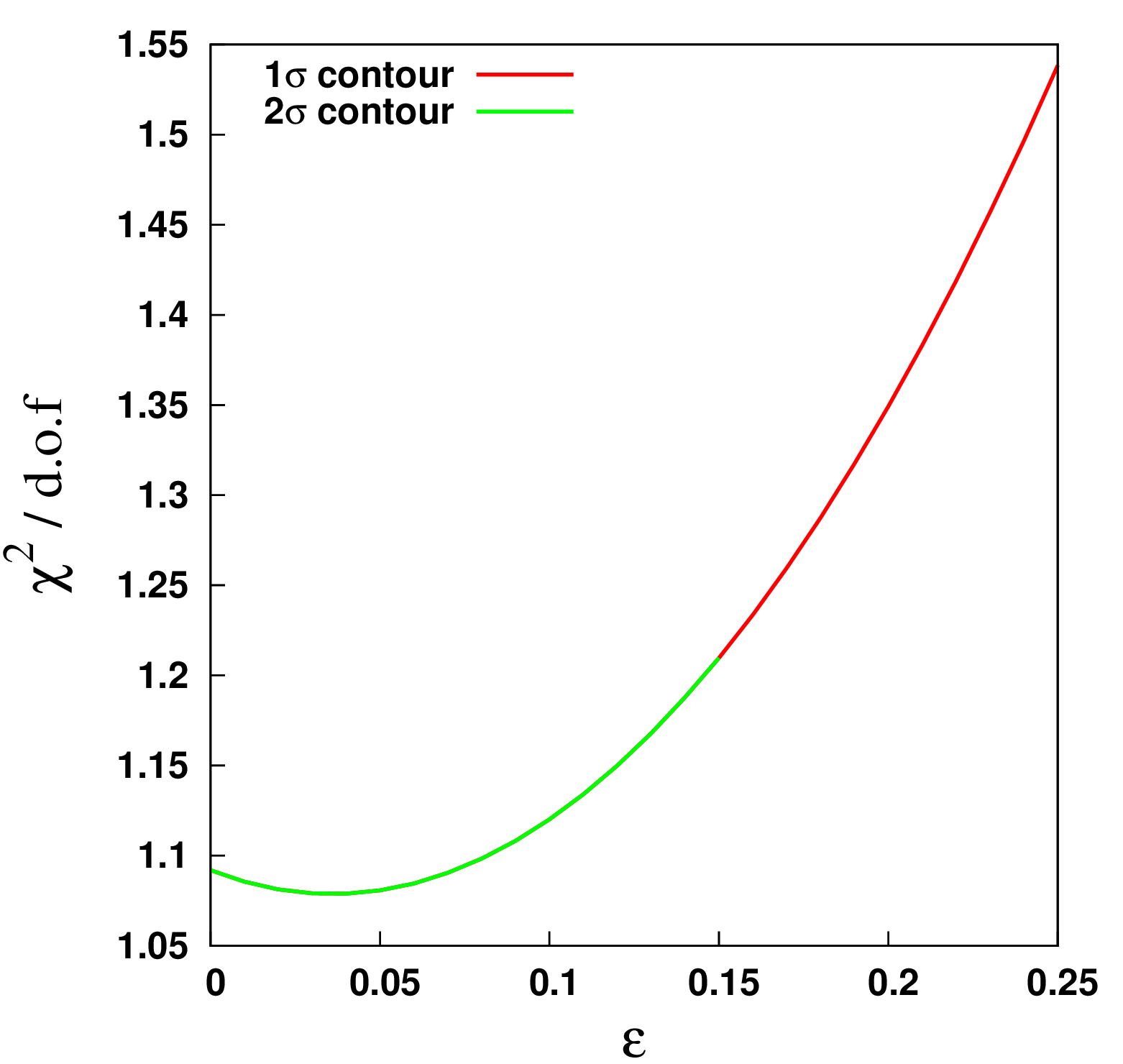} ~~ \includegraphics[height=5cm,width=7cm]{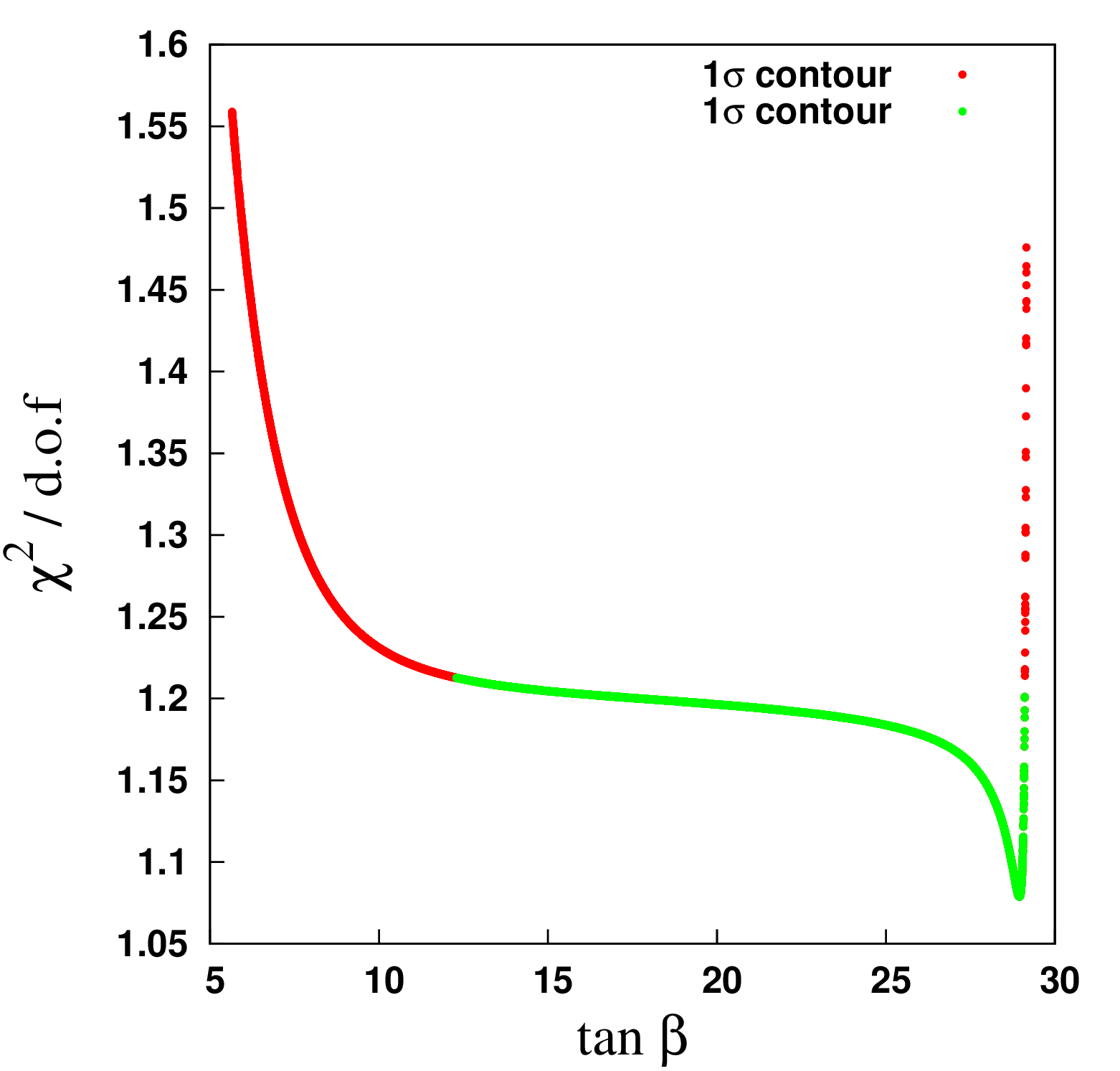}
\caption{$1\sigma$ and $2\sigma$ contours for $\varepsilon$ and $\tan\beta$ from $\chi^2$ minimization obtained for BP3. 
The green line indicates $1\sigma$ reach and the red line indicates $2\sigma$ reach of the parameters.}
\label{fig:BP3_eps_tanb}
\end{figure}

We have varied $\tan\beta$ between 2 and 50, and $\varepsilon$ between
0 and 0.7. Note that large $\tan\beta\gsim 50$ is disfavored by the recent LHCb results on $B_s\to \mu^+\mu^-$, and very low $\tan\beta\lsim 2$ are usually not considered due to radiative electroweak symmetry breaking arguments. The $1\sigma$ and $2\sigma$ contours for $\tan\beta$ and $\varepsilon$ for various `snapshot' values of
the high-scale parameters are presented in Figures \ref{fig:BP1_eps_tanb}, \ref{fig:BP2_eps_tanb} and \ref{fig:BP3_eps_tanb}. 
We have chosen the benchmark points for signal prediction, ensuring that we stay within $2\sigma$ for both of the fitted parameters 
($\varepsilon$ and $\tan\beta$). 
\begin{figure}[h!]
\begin{center}
\includegraphics[height=5cm,width=7cm]{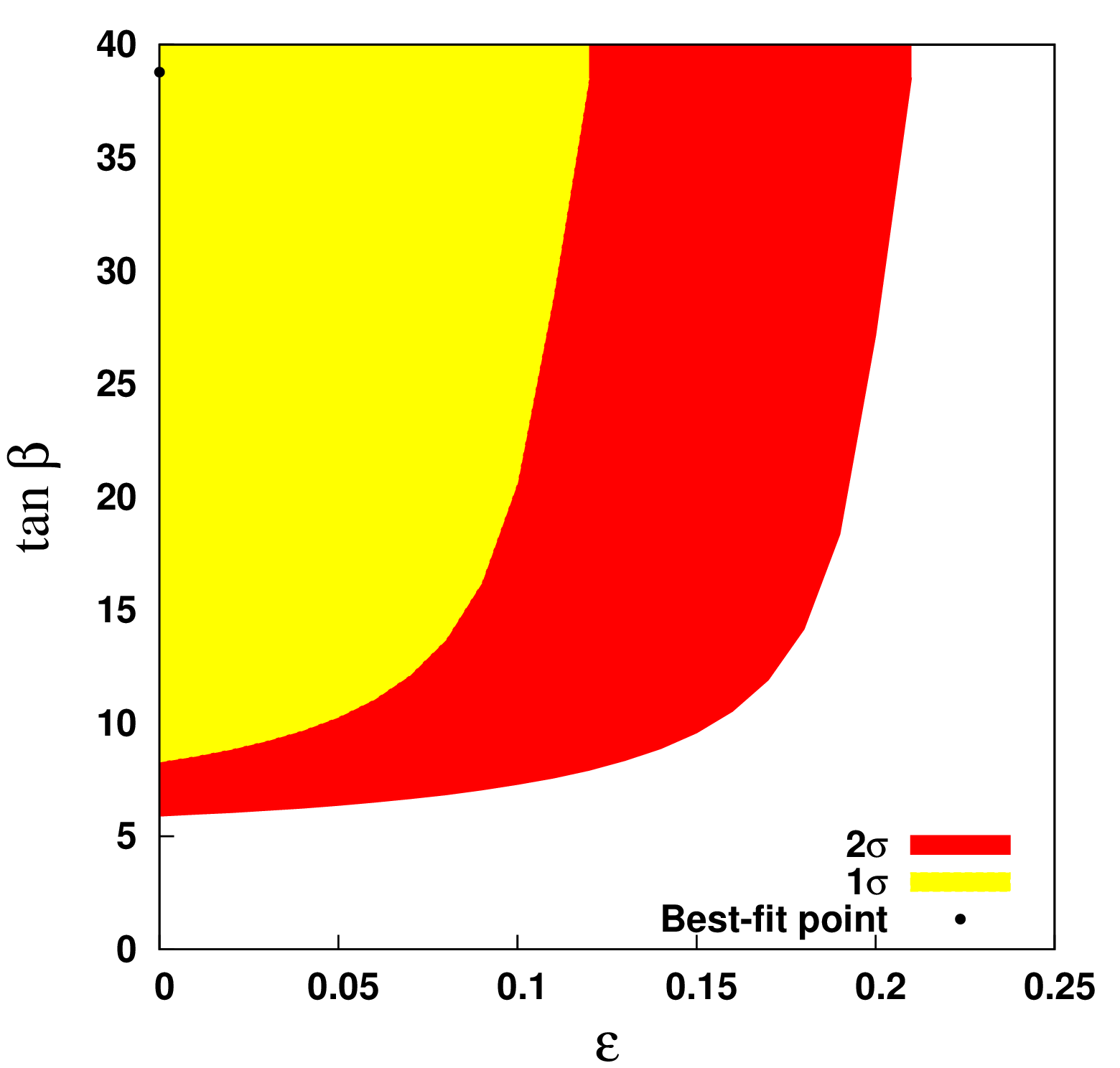}~~
\includegraphics[height=5cm,width=7cm]{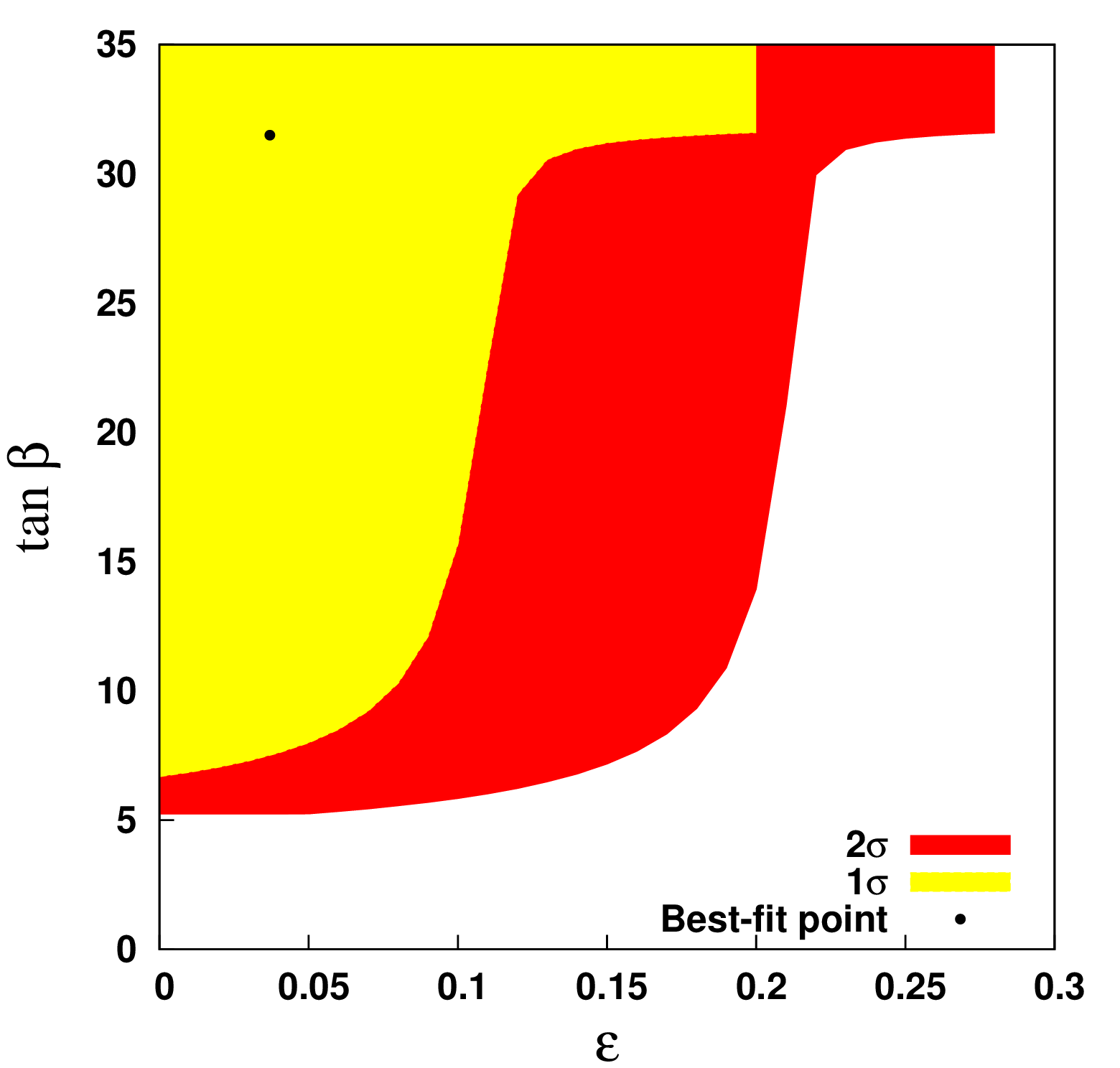}\\
\includegraphics[height=5cm,width=7cm]{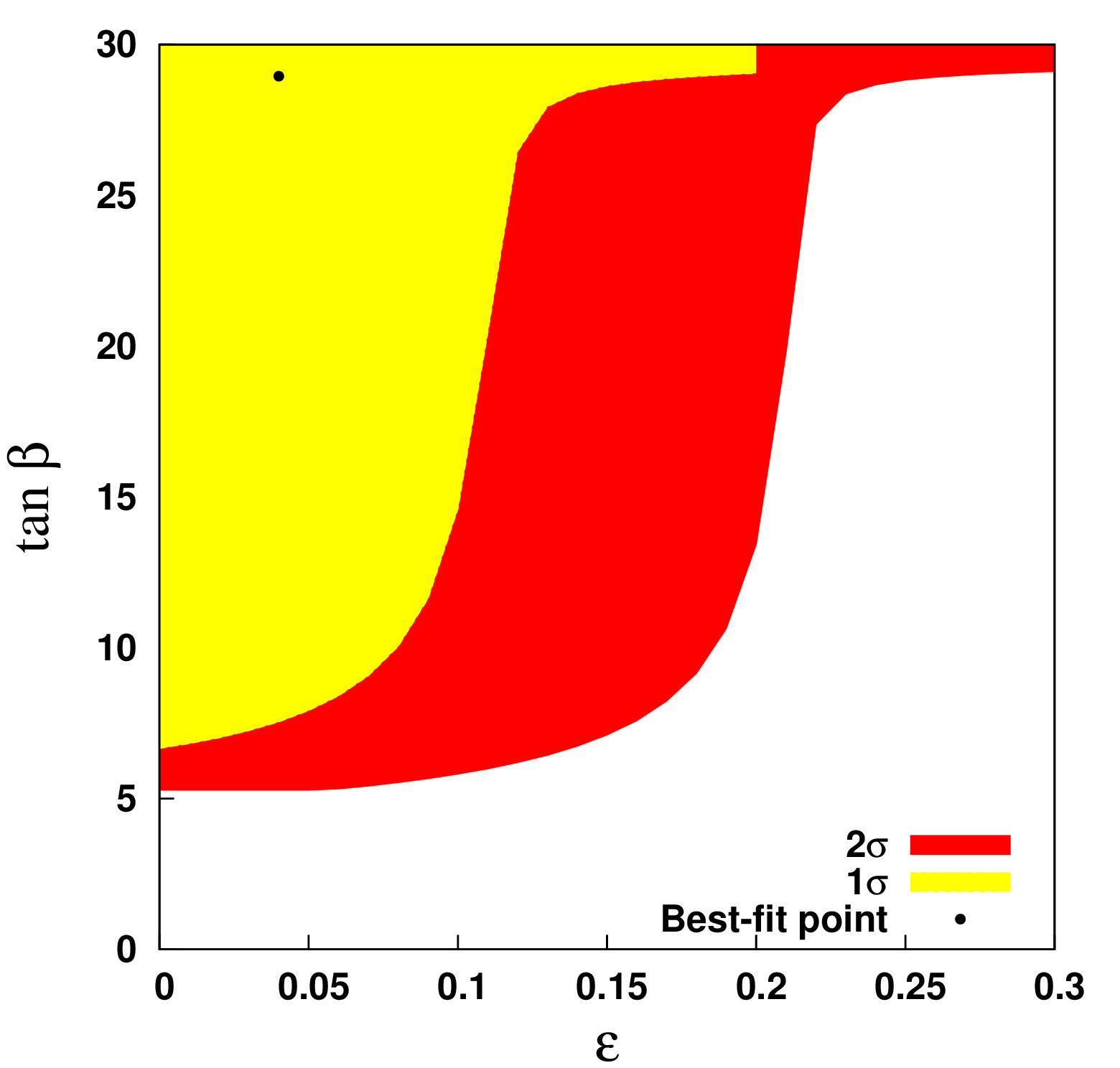}
\end{center}
\caption{$1\sigma$ and $2\sigma$ allowed contours in the $\varepsilon$ - $\tan\beta$ plane from global analysis 
of the Higgs data for the three benchmark points in our model. The black dots indicate the best fit values of 
 $\tan\beta$ and $\varepsilon$ obtained from our analysis. }  
\label{fig:cont}
\end{figure}

As manifested in these contour plots, the minimum $\chi^2$-value is obtained for $\varepsilon=0, 0.037$ and $0.04$ for BP1, BP2 and BP3 respectively 
and intermediate values of $\tan\beta$ 
around 30-40 for all the benchmark points. Also, there exists an upper limit on $\varepsilon$ to be consistent with the LHC Higgs data. 
The 68.27\% ($\Delta \chi^2$ = 1) and 95.45\% ($\Delta \chi^2$ = 4) CL limits derived from Figures~\ref{fig:BP1_eps_tanb}-\ref{fig:BP3_eps_tanb} are summarized in Table~\ref{tab:limits}, and also shown in Figure~\ref{fig:cont}. 
These limits are comparable to those obtained in a recent model-independent global fit~\cite{Giardino:2013bma}, and much stronger than the 
direct search limits from associated production of Higgs with $Z$~\cite{ATLAS-011, cmsinv} as well as those derived from monojet 
searches~\cite{Bai:2011wz}. 
\begin{table}[h!]
\begin{center}
\begin{tabular}{||c||c|c||c|c||c|c||}\hline\hline
Parameter & \multicolumn{2}{c||}{BP1} & \multicolumn{2}{c||}{BP2} & \multicolumn{2}{c||}{BP3} \\ \cline{2-7}
& $1\sigma$ & $2\sigma$ & $1\sigma$ & $2\sigma$ & $1\sigma$ & $2\sigma$ \\
\hline\hline
$\varepsilon$ & $<0.07$ & $<0.16$ & $<0.15$ & $<0.24$ & $<0.15$ & $<0.25$ \\
$\tan\beta$ & 12.0-38.8 & 6.9-38.8 & 14.5-31.6 & 5.7-31.6 &  12.3-29.1 & 5.6-29.2\\
\hline\hline 
\end{tabular}
\end{center}
\caption{The $1\sigma$ and $2\sigma$ limits on the invisible Higgs BR and the MSSM $\tan\beta$ parameter obtained from the marginalized 
plots (Figures~\ref{fig:BP1_eps_tanb}-\ref{fig:BP3_eps_tanb}) for the chosen benchmark points in SISM. }
\label{tab:limits}
\end{table}
\subsection{Upper limit on the Dirac Yukawa Coupling}
\begin{figure}[h!]
\centering
\includegraphics[height=5cm,width=7cm]{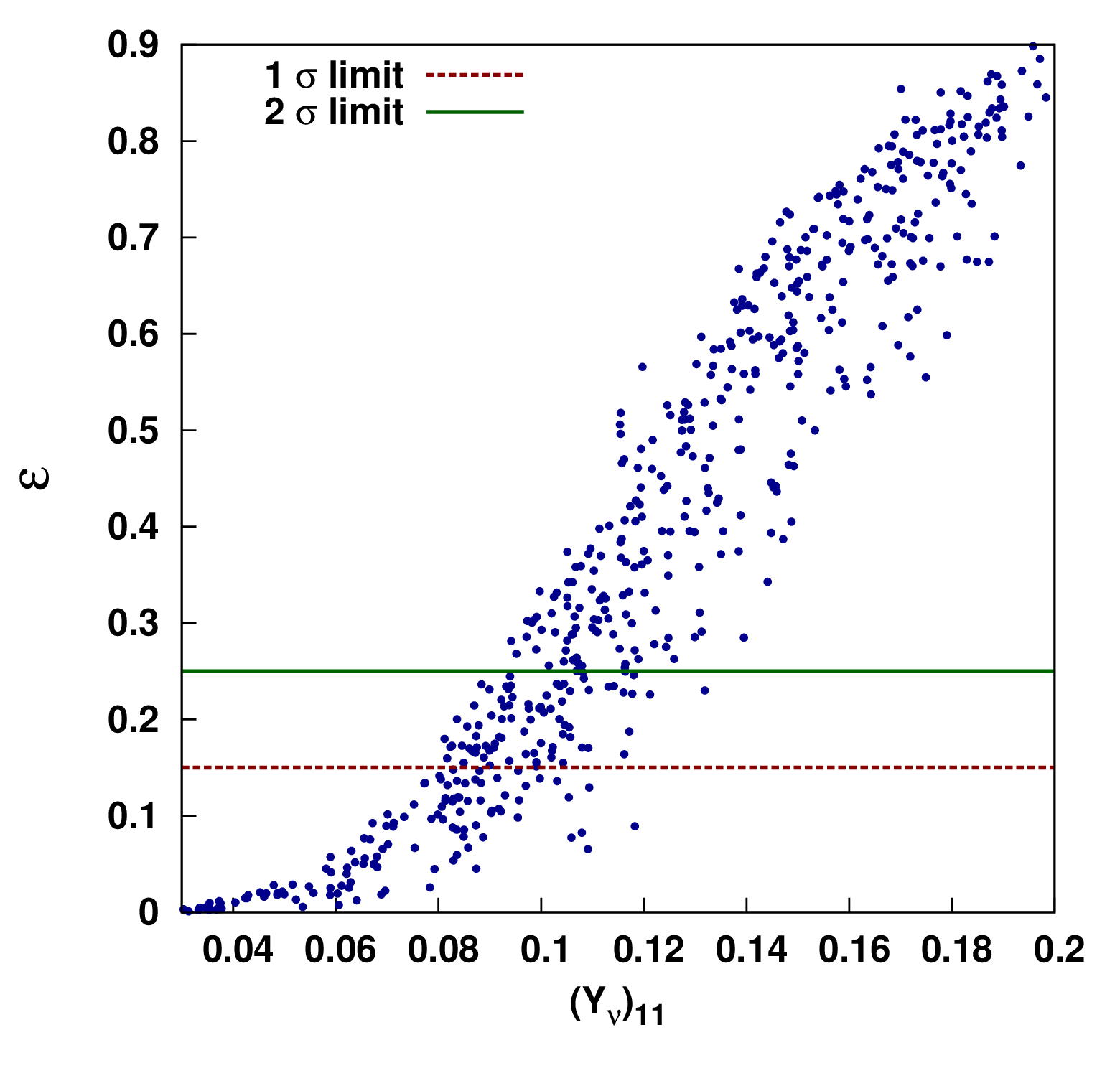}
\caption{The invisible Higgs branching fraction as a function of the Dirac Yukawa coupling. The $1\sigma$ and $2\sigma$ upper limits on the invisible branching fraction derived earlier are also shown. }
\label{fig:inv_br}
\end{figure}
An upper limit on the invisible Higgs branching ratio, as derived in Table~\ref{tab:limits} from a global analysis of the LHC Higgs data, will put an upper limit on the magnitude of the Dirac Yukawa coupling in the model. To illustrate this, we show in Figure~\ref{fig:inv_br} the variation of the invisible Higgs branching fraction as a function of the Yukawa parameter, $(y_{\nu})_{11}$.
This plot is obtained for a fixed $A_0 (\sim -2.8~{\rm TeV})$ and fixed $B_{M_R}$ and $B_{\mu_S}$ as given below Eq.~(\ref{range}). However, other parameters are varied in the ranges mentioned in Eq.~(\ref{range}). Note that, the invisible higgs BR is insensitive to other entries of $y_{\nu}$. We obtain a spread of the points as during the scan the Higgs mass fluctuates a
little bit around its central value. Also the $M_R$ parameters vary which means that the LSP mass is not fixed at a particular value. It roughly varies between 20 - 62 ${\rm GeV}$, and for most of the points, lie in the 30 - 62 ${\rm GeV}$ range.  
As can be seen from the plot, the invisible Higgs branching fraction roughly grows with the Yukawa coupling 
in the kinematically allowed region. Thus an upper limit on the Dirac Yukawa coupling in the model follows from the upper limit on $\varepsilon$, as can be read off from the $1\sigma$ and $2\sigma$ lines in 
Figure~\ref{fig:inv_br}. Note that the upper limit of order of 0.10 on $y_\nu$ derived from this analysis is stronger than those derived from the Higgs visible decay~\cite{BhupalDev:2012zg} for a heavy neutrino mass larger than the Higgs mass. Comparable limits on $y_{\nu}$ in similar TeV scale seesaw models are obtained from charged-lepton flavour violating decays for the range of heavy neutrino masses we have considered here~\cite{Ibarra:2011xn,Dinh:2012bp}.

\begin{figure}[h!]
\centering
\includegraphics[height=5cm,width=7cm]{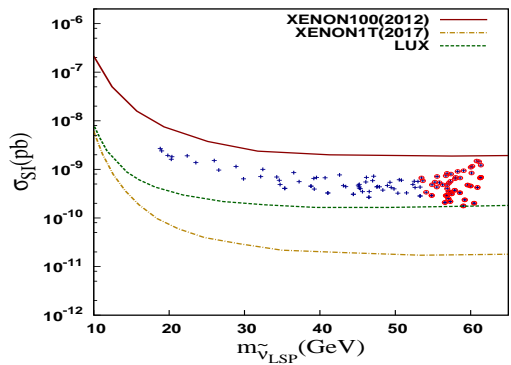}
\caption{Spin-independent cross section as a function of the sneutrino LSP mass for the points satisfying the 
$2\sigma$ upper limit on the invisible Higgs BR. The circled points also satisfy the relic density constraint.}
\label{fig:sigma_si2}
\end{figure}
The bound on $\varepsilon$ also constrains the allowed parameter space for the DM-nucleon elastic 
scattering cross section in this model. This is shown in  Figure~\ref{fig:sigma_si2} which is basically 
a zoomed-in version of Figure~\ref{fig:sigma_si} focusing on the light DM region and with only those points obeying the $2\sigma (< 25\%)$ limit on $\varepsilon$. As can be seen from the plot, all these points are just 
below the current sensitivity of XENON100 experiment, but can be {\it completely} probed by the future 
experiments such as LUX and XENON1T.
\subsection{Some benchmark points}
\begin{table}[h!]
\begin{center}
\begin{tabular}{||c|c|c|c||}\hline\hline
Input parameter & BP1 & BP2 & BP3 \\ 
\hline\hline
$\tan\beta$ & 25 &20 & 25\\ 
$y_\nu$  & (0.095,0.090,0.090) & (0.074,0.064,0.064)& (0.0701,0.010,0.010)\\
$M_R$ (GeV) & (192.7,1000,1000)& (679.16,1000,1000)& (798,1000,1000)\\
$B_{\mu_S}$ (GeV$^2$) & $10^{-4}$ & $10^{-4}$ & $10^{-4}$\\
$B_{M_R}$ (GeV$^2$) & $10^6$ & $10^6$ & $10^6$\\
$\mu_S$ (eV) & {\scriptsize $\left(\begin{array}{ccc}
0.55 & 6.06 & 1.92\\
6.06 & 107.86 & 87.97\\
1.92 & 87.97 & 116.73\\
\end{array}\right)$} & {\scriptsize $\left(\begin{array}{ccc}
11.25 & 38.73 & 12.29\\
38.73 & 213.51 & 174.15 \\
12.29 & 174.15 & 231.08
\end{array}\right)$} & {\scriptsize $\left(\begin{array}{ccc}
17.30 & 307.19 & 97.46\\
307.19 & 8738.55 & 7127.59\\
97.46 & 7127.59 & 9457.73
\end{array}\right)$} \\
\hline\hline
\end{tabular}
\end{center}
\caption{Benchmark values of $\tan\beta$ and the low scale neutrino sector parameters for the chosen 
benchmark points in Table~\ref{tab:mssm_scan}. }
\label{tab:input}
\end{table}
In Table~\ref{tab:input} we present some benchmark values of the remaining model parameters not shown in 
Table~\ref{tab:mssm_scan} as allowed by the invisible Higgs decay constraints. We have chosen the neutrino sector 
parameters to be diagonal, except for $\mu_S$ which was fixed by fitting the central global fit values of the neutrino 
oscillation parameters given in Eq.~(\ref{eq:oscil}). For illustration, we have assumed a normal hierarchy of neutrino 
masses with $m_1=10^{-5}$ eV and the Dirac $CP$ phase $\delta=0$ in the PMNS matrix. It is clear from the choice of mSUGRA 
parameters in Table~\ref{tab:mssm_scan} that our low-energy MSSM particle spectrum is consistent with the current limits 
from direct SUSY searches~\cite{atlas-susy, cms-susy}. We also calculate the other low-energy observables in the flavor 
sector using {\tt SPheno} and in the DM sector using {\tt micrOMEGAs} for 
the particle spectrum generated from {\tt SPheno} using the input values shown in Tables~\ref{tab:mssm_scan} and \ref{tab:input}. 
These results, summarized in Table~\ref{tab:low}, ensure that the chosen 
benchmark points are consistent with all the existing collider, cosmological and low energy constraints 
listed below (within their $3\sigma$ allowed range, where applicable):
 (i) $m_h=125\pm 2$ GeV~\cite{CMS1,ATLAS3}, (ii) $\Omega h^2=0.1199\pm 0.0027$~\cite{Ade:2013zuv}, 
 (iii) $\sigma_{\rm SI}<5\times 10^{-9}$ pb for $m_{\rm DM}\simeq 50$ - 60 GeV~\cite{Aprile:2012nq}, 
 (iv) $\delta a_\mu=(26.1\pm 8.0)\times 10^{-10}$~\cite{Hagiwara:2011af} and $\delta a_e=(109\pm 83)\times 10^{-14}$~\cite{Aoyama:2012wj}, 
 (v) BR$(B\to X_s\gamma)=(3.21\pm 0.33)\times 10^{-4}$~\cite{Lees:2012ym}, 
 (vi) BR$(B_s\to \mu^+\mu^-)= \left(3.2^{+1.5}_{-1.2}\right)\times 10^{-9}$~\cite{Aaij:2012nna}, 
 (vii) constraints from the LFV decays~\cite{PDG}, and (viii) non-unitarity constraints in the neutrino sector~\cite{Abada:2007ux, Antusch:2008tz}.

\begin{table}[h!]
\begin{center}
\begin{tabular}{||c|c|c|c||} \hline\hline
Parameter & BP1 & BP2 & BP3 \\ \hline\hline
$m_h$ (GeV) & 124.69 & 125.79 & 125.78 \\ \hline
$\Omega_{\rm DM}h^2$ &  0.114 & 0.122 & 0.112 \\ 
$\sigma_{\rm SI}$ (pb) & $3.38\times 10^{-10}$ & $5.26\times 10^{-10}$ & $5.56\times 10^{-10}$ \\ 
\hline
$\delta a_\mu$ & $3.1\times 10^{-10}$ & $2.5\times 10^{-10}$ & $3.4\times 10^{-10}$ \\
$\delta a_e$ & $7.0\times 10^{-15}$ & $5.7\times 10^{-15}$ & $7.8\times 10^{-15}$ \\ \hline
BR$(B\to X_s\gamma)$ & $2.9\times 10^{-4}$ & $3.1\times 10^{-4}$ & $3.1\times 10^{-4}$ \\
BR$(B_s\to \mu^+\mu^-)$ & $3.7\times 10^{-9}$ & $3.5\times 10^{-9}$ & $3.6\times 10^{-9}$ \\ \hline
BR$(\mu\to e\gamma)$ & $5.2\times 10^{-22}$ & $1.1\times 10^{-22}$ & $3.5\times 10^{-22}$ \\
BR$(\tau\to e\gamma)$ & $9.8\times 10^{-21}$ & $2.1\times 10^{-21}$ & $6.6\times 10^{-21}$\\
BR$(\tau\to \mu\gamma)$ & $1.6\times 10^{-16}$ & $3.5\times 10^{-17}$ & $1.1\times 10^{-16}$ \\
BR$(\mu\to 3e)$ & $1.1\times 10^{-22}$ & $8.9\times 10^{-25}$ & $2.7\times 10^{-24}$\\
BR$(\tau\to 3e)$ & $6.8\times 10^{-22}$ & $2.5\times 10^{-23}$ & $7.7\times 10^{-23}$\\
BR$(\tau\to 3\mu)$ & $2.8\times 10^{-16}$ & $3.0\times 10^{-19}$ & $7.9\times 10^{-19}$\\ \hline
$|\eta_{ee}|$ & $3.67\times 10^{-3}$ & $1.79\times 10^{-4}$ & $1.16\times 10^{-4}$\\
$|\eta_{\mu\mu}|$ & $1.22\times 10^{-4}$ & $6.18\times 10^{-5}$ & $1.51\times 10^{-6}$\\
$|\eta_{\tau\tau}|$ & $1.22\times 10^{-4}$ & $6.18\times 10^{-5}$ & $1.51\times 10^{-6}$\\ \hline\hline
\end{tabular}
\end{center}
\caption{The Higgs mass, relic density, spin-independent cross section, anomalous magnetic moments and the relevant low-energy flavor sector observables in the SISM for the three chosen BPs.}
\label{tab:low}
\end{table}
\section{Collider Analysis}
The possibility of an invisible Higgs signature at the LHC has been explored both theoretically~\cite{Frederiksen:1994me, Eboli:2000ze, Godbole:2003it, Davoudiasl:2004aj, Zhu:2005hv, Bai:2011wz, Ghosh:2012ep} and experimentally~\cite{ATLAS-011, cmsinv, Bansal:2010zz}. These studies show that the most promising Higgs production channel for detecting an invisibly decaying Higgs is the vector boson fusion (VBF), and the next promising channel is its associated production with $Z$. 
In the VBF channel, Higgs is produced from vector bosons originated by radiation off two initial 
state quarks along with two jets, and subsequently decays into invisible final states: 
$pp\to qqh \to qq+\mET$. Thus the final state consists of two jets widely separated in rapidity together with large missing transverse energy.  In the $Zh$ associated production channel, 
the $Z$ decays into two 
oppositely charged leptons and the Higgs decays invisibly: 
$q\bar{q}\to Z+h\to \ell^+\ell^-+\mET$. Note that the leptonic decay channel of $Z$ is known to be cleaner than its hadronic counterpart with $b$-jets. 
One can also look for an associated $Wh$ production where $W$ decays leptonically to give rise to 
a $\ell+\mET$ final state. However, the signal acceptance efficiency in this channel is found to 
be very small, and hence, the corresponding exclusion limit is much worse than that from the $Zh$ channel~\cite{Gagnon:sna}. 

In addition to these channels, the dominant Higgs production channel at the LHC, namely, gluon-gluon fusion (ggF), can give rise to a monojet+large $\mET$ signal with the jet coming from initial 
state radiation and Higgs decaying invisibly. But the QCD background for this process is too 
large, and moreover, it is hard to isolate the new physics effects only for the Higgs 
invisible decay since these effects could also show up in loops to modify the ggF production 
cross section. The $\sqrt s=7$ TeV search results in this channel~\cite{ATLAS-monoj, CMS-monoj} 
were translated to a weak upper limit on $\varepsilon<$ 0.4 - 0.6~\cite{Bai:2011wz} depending on the jet $p_T$ threshold selection. 
Finally, the other relevant Higgs production channel, 
namely in association with top pairs, has a much smaller cross section~\cite{Dittmaier:2011ti}, and involves complex final states which require a very sophisticated analysis. Therefore, 
we will focus on the VBF channel 
with 2 jets+$\mET$ final states and the $Zh$ channel with $\ell^+\ell^-+\mET$ final states for the collider analysis of invisible Higgs signature in our model. We show our analysis results for $\sqrt s=14$ TeV LHC.  

\subsection{Event generation}
The SUSY particle spectrum and various decay branching fractions in our model have been 
calculated using {\tt SPheno}~\cite{spheno}. The {\tt SLHA} files are then fed to {\tt PYTHIA}
(version 6.409)~\cite{pythia} for event generation. The initial and final state radiation of quarks and gluons, multiple interactions, decay, hadronization, fragmentation and jet formation are 
implemented following the standard procedures in {\tt PYTHIA}. 
The factorization and renormalization scales $\mu_R$ and $\mu_F$ respectively are both set at 
the parton-level center of mass energy $\sqrt{\hat s}$. We have used the {\tt CTEQ5L}~\cite{cteq} 
parton distribution functions in our analysis. The jets with $p_T > 20~{\rm GeV}$ and $|\eta | < 4.5$ have been constructed using the cone algorithm via {\tt PYCELL}. To simulate detector effects, 
we take into account the smearing of jet 
energies by a Gaussian probability density function of width $\sigma (E)/E_j = (0.6/\sqrt{E_j[{\rm GeV}]}) + 0.03$, $E_j$ being the unsmeared jet energy~\cite{Barr}. 

Following are the selection cuts that we have used to find the final state leptons and
jets:
\begin{itemize}
\item For final state electrons and muons we use $p_T > 15~{\rm GeV}$ and $p_T > 10~{\rm GeV}$ respectively. For both, we take $|\eta| < 2.4$.
\item Lepton-lepton separation $\Delta R_{\ell\ell} > 0.2$, where $\Delta R = \sqrt{(\Delta\eta)^2 + (\Delta\phi)^2}$.
\item Lepton-jet separation $\Delta R_{\ell j} > 0.4$.
\item Scalar sum of $E_T$ deposits by hadrons within a cone of $\Delta R \le 0.2$ around a lepton
must be less than $0.2 p_T^\ell$ to ensure lepton isolation.
\item Jet-jet separation $\Delta R_{jj} > 0.4$.
\end{itemize}

Depending on the hadronic or leptonic signal final states, we use specialized 
selection criteria, as discussed below. 
\subsection{The VBF channel}
In this case, the two leading high $p_T$ jets in the final state are produced in forward and 
backward directions with rapidities opposite in sign and widely separated. Also due to the 
invisible decay of the Higgs, one expects a large amount of missing energy. These features 
largely help to reduce the SM background. The dominant SM background for this signal can come 
from:\\  
(i) $W$+ jets, where $W$ decays leptonically and the lepton escapes detection. \\
(ii) $Z$ + jets, where $Z$ decays into two neutrinos. \\
(iii) mismeasured QCD events giving fake missing energy. \\
The contributions from non-VBF processes, for instance, from hard QCD production of a single 
Higgs or a Higgs with associated quarks and gluons, must also be taken into account for the 
signal. Despite its poor efficiency to pass the background reducing cuts, due to its large production 
cross section the ggF channel can contribute 4-5\% of the VBF signal~\cite{DelDuca:2001fn, Nikitenko:2007it}. 
The following cuts have been used to reduce the background: 
\begin{itemize}
 \item Absolute rapidity difference between the two leading jets, 
$|\eta_{j_1} - \eta_{j_2}| > 4.0$. To ensure that the two jets are produced in forward and backward directions, we require $\eta_{j_1}\cdot \eta_{j_2} < 0$.
 \item A jet veto with $p_T > 40$ GeV in the central region since we don't expect any jets in the 
rapidity gap of the two jets for a pure VBF process. We discard jets with $|\eta| < 2.5$   
 \item Invariant mass of the two leading jets, $M_{jj}> 1.8~{\rm TeV}$ .
 \item A $\mET$ cut of $100~{\rm GeV}$.
\end{itemize}
The $\mET$ and $M_{jj}$ cuts reduce the background efficiently, and also reduce the QCD 
contributions significantly. We note here that two additional cuts have been 
occasionally used in the literature for isolating events with invisible final states. These are $\Delta\phi(j,\mET)$ and $\Delta\phi(j_1,j_2)$. We have checked that these cuts reduce the signal cross section to far too a level in our case. Therefore, we have dropped them and used the optimal set of event selection criteria mentioned above.  

The cross sections for the signal corresponding to the benchmark points chosen earlier as well as dominant backgrounds 
coming from $W + n$-jets and $Z + n$-jets ($n=0,1,2,3$) are shown in 
Table~\ref{tab:vbf}. The background events were generated using {\tt Alpgen}~\cite{Mangano:2002ea} 
at the partonic level and then passed to {\tt PYTHIA} for showering. While interfacing, we have 
incorporated the 
{\tt MLM}~\cite{MLM} prescription to match between the hard jets generated by 
{\tt Alpgen} and the soft radiation jets generated by {\tt PYTHIA} in order to avoid double 
counting. Since the background channels have huge inclusive cross sections, we generated 
at least $\sim 10^7$ unweighted events for all the channels in {\tt Alpgen} in order to get proper 
convergence.  For the signal cross section, we show the values obtained for VBF as well as for 
other hard processes $gg\rightarrow h$, 
$q\bar q\rightarrow gh$, $qg\rightarrow qh$ and $gg\rightarrow gh$.   
\begin{table}[h!]
\begin{center}
\begin{tabular}{| c | c | c|} \hline 
Channel & Production cross section (pb) & Cross section after cuts (fb)\\ \hline
BP1 (VBF) & 3.76 & 0.99  \\ 
BP1 (others) & 125.9 & 0.16 \\ 
BP2 (VBF) & 3.72 & 1.55 \\
BP2 (others) & 125.4 & 0.25 \\
BP3 (VBF) & 3.73 & 1.72  \\
BP3 (others) & 125.7 & 0.25 \\ \hline
$W$ + n-jets & 56848.54  & 46.57  \\
$Z$ + n-jet & 10198.72 & 24.90  \\ \hline
\end{tabular}
\end{center}
\caption{Final cross sections obtained for all the signal and SM background channels for a $14~{\rm TeV}$ LHC run. For the background channels, 
the cross sections in the 2nd column are those of the final states, i.e, $W$ and $Z$ decays into 
lepton-neutrino and 
two neutrino channels respectively. The cross sections in the 3rd column are the ones obtained 
after all the selection and background reduction cuts. n-jets corresponds to 0, 1, 2, 3 jets combined result.}
\label{tab:vbf}
\end{table}
It is clear that, despite the large production cross section, contributions to the signal coming 
from channels other than VBF channel are very small after applying all the cuts. Also the SM backgrounds are hugely suppressed after all the cuts, optimized for a good signal significance, 
$\frac{S}{\sqrt{S+B}}$, where $S$ and $B$ stand for the signal and background strengths 
respectively. 
From Table~\ref{tab:vbf}, we find that for BP1 with the maximum (2$\sigma$ allowed) invisible 
branching ratio $\varepsilon_{\rm max}=0.16$ for the Higgs, we obtain a $3\sigma$ signal 
significance at $500~{\rm fb}^{-1}$ whereas for BP2 and BP3 with $\varepsilon_{\rm max}=0.24$ and 
$0.25$ respectively, we can obtain a $3\sigma$ significance at $200~{\rm fb}^{-1}$.

\subsection{The $Zh$ channel}
In this channel, we are interested in the leptonic decay of $Z$ leading to a same-flavor, 
opposite-sign dilepton plus large missing energy from the invisible decay of the Higgs.   
 The dominant SM background in this case comes from:\\
(i) $WW$ production, where both the $W$'s decay leptonically. \\
(ii)$WZ$ production, where $Z$ decays into two charged leptons and $W$ into a charged lepton and 
neutrino, and one charged lepton misses detection. \\
(iii)$ZZ$, where one $Z$ decays into two charged leptons and the other into two neutrinos. \\
(iv)$t\bar t$ production followed by $t\to Wb$, where both the $W$'s decay leptonically and the 
$b$-jets escape detection. 

We use the following cuts to reduce the SM background:
\begin{itemize}
\item A jet veto with $p_T > 20~{\rm GeV}$ and $|\eta| < 4.5$ since the signal consists of no 
jets.
\item Dilepton invariant mass $|M_Z-M_{\ell\bar{\ell}}|<10$ GeV since the two charged  
leptons in the final state come from $Z$-boson decay. 
\item Di-lepton transverse mass $M_T^{\ell\ell} \ge 150~{\rm GeV}$, where 
$M_T^{\ell\ell} = \sqrt{p_T^{\ell\ell}\mET [1 - cos\phi(p_T^{\ell\ell},\mET)]}$. This is because 
the $Z$-boson and the Higgs are more likely produced back-to-back for the signal, thus leading to 
a harder transverse mass distribution for the di-lepton system, as can be seen from 
Figure~ \ref{fig:mtll}. 
\begin{figure}[h!]
\centering
\includegraphics[width=7cm]{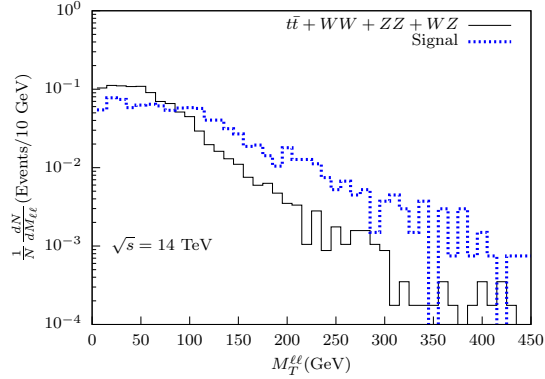}
\caption{Normalized transverse mass distribution for the di-lepton system in the $Zh$ signal and combined SM background events at $14~{\rm TeV}$ LHC.} 
\label{fig:mtll}
\end{figure}
\item $\mET>100$ GeV since the signal is expected to have a harder $\mET$ distribution, as 
verified by Figure~\ref{fig:met}. 
\begin{figure}[h!]
\centering
\includegraphics[width=7cm]{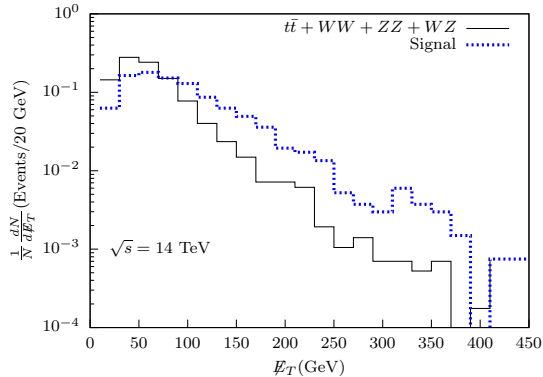}
\caption{Missing transverse momentum distribution of the $Zh$ signal and combined SM background events at $14~{\rm TeV}$ LHC.}
\label{fig:met}
\end{figure}
\end{itemize} 

Table~\ref{tab:signal} shows the production cross sections and final cross sections after all the cuts for the signal corresponding to the chosen benchmark point as well as for the SM background. 
\begin{table}[h!]
\begin{center}
\begin{tabular}{| c | c | c|} \hline 
Channel & Production cross section (pb) & Cross section after cuts (fb)\\ \hline
BP1 & 0.53 & 0.35  \\ 
BP2 & 0.51 & 0.51  \\ 
BP3 & 0.51 & 0.52  \\ \hline
$WW$ & 76.51 & 0.38 \\
$ZZ$ & 10.58 & 7.77 \\
$WZ$ & 28.95 & 8.83 \\
$t\bar t$ & 370.20 & 0.92 \\ \hline
\end{tabular}
\end{center}
\caption{Cross sections obtained for all the signal and SM background channels for $14~{\rm TeV}$ 
LHC. The 2nd column shows the production 
cross sections for various channels and the 3rd column after all the selection and background 
reduction cuts.}
\label{tab:signal}
\end{table}
As can be seen from Table~\ref{tab:signal}, this channel has a huge SM background which can 
easily dominate over the signal events. The signal significance factor is quite low in this case 
for all the benchmark points. For BP2 and BP3 with a maximum invisible BR of Higgs $\varepsilon \sim 0.25$,
the signal can achieve a significance of $3\sigma$ only at $600~{\rm fb}^{-1}$ luminosity, 
whereas for BP1, to get such significance, we need to go beyond $1300~{\rm fb}^{-1}$ 
at $14~{\rm TeV}$ center of mass energy. 

The reason for better LHC detection prospects for BP2 and BP3 compared to BP1 can be understood by comparing their 
corresponding particle spectra. The invisible higgs branching ratio 
$\varepsilon$ depends on the masses of the Higgs, LSP sneutrino and the Higgs-sneutrino-sneutrino coupling. 
Since the masses of the parent and daughter particles are almost identical for all the 
three cases, what makes the difference in the invisible decay width is the coupling which depends on the 
amount of mixing of the singlet sneutrinos with the left-handed ones. 
The singlet components dominate the lightest sneutrino mass eigenstates for all the three benchmarks because of 
the large $B_{M_R}$ term in the off-diagonal of the sneutrino mass matrix given by Eq.~(\ref{eq:svmass}). This 
parameter does not change for the three benchmark points and as a result, the right-handed components are not expected 
to vary much from BP1 to BP3. However, these components also depend on the matrices 
$m_N^2$, $m_S^2$ and $M_R$. Here $m_N^2$ and $m_S^2$ scale as $m_0^2$. Now from BP1 to BP3, $m_0$ keeps decreasing 
and $M_R$ keeps increasing. Hence the diagonal terms in Eq.~(\ref{eq:svmass}), although comparable, keep 
increasing slightly. This brings down the right-handed contribution in the lightest state by a very small 
amount from BP1 to BP3 (to be precise, the component comes down from 0.716 to 0.710). 
On the other hand, as the absolute value of the trilinear term $A_{\nu}$ in Eq.~(\ref{eq:svmass}) 
decreases from BP1 to BP3, it brings down the left-component and increases the right-handed component. 
As a result of these competing effects, the left-component of the sneutrino LSP, and hence, the Higgs invisible decay width 
increases from BP1 to BP3, thus enhancing the LHC detection prospects. 

Before concluding this section, we wish to emphasize an important distinction of our scenario from similar signals in 
the MSSM with a neutralino LSP which could otherwise obliterate the distinct collider signals of our model. 
As already pointed out in~\cite{BhupalDev:2012ru}, the pure cMSSM case can be 
distinguished from the SISM case by studying the same-sign dilepton+jets+$\mET$~ signal which is enhanced in the SISM case. 
Also the SISM case has a much harder $\mET$ tail compared to the cMSSM case which can be used as another distinguishing 
feature of our model. Finally, the ``residual MSSM backgrounds'' can be reduced/removed by studying the effective mass 
distribution of the events, defined as the scalar sum of the lepton and jet transverse momenta and missing transverse energy:
\begin{eqnarray}
	M_{\rm eff} = \sum|p_T^{\ell}|+\sum|p_T^j|+\mET .
\end{eqnarray}
Taking into account the current limits on the sparticle masses, the $M_{\rm eff}$ distribution of events arising from sparticle production will be considerably harder in the pure MSSM case than in our case. Note that the cascade decays 
involving chargions can also be used to measure the mass of the sneutrino LSP at the LHC applying the $m_{T_2}$ endpoint technique~\cite{Belanger:2011ny}.   
\section{Conclusion}
We have shown that supplementing the cMSSM framework with inverse seesaw mechanism for 
neutrino masses can give rise to a light sneutrino DM candidate with mass around $50~{\rm GeV}$ 
while being consistent with all the existing collider, cosmological as well as low-energy 
constraints. Such a light scalar DM also leads to the possibility of the lightest $CP$ even Higgs boson in the MSSM decaying invisibly into two such DM particles induced by a soft trilinear 
 coupling.  We have explored this possibility in details by performing a  
global $\chi^2$-analysis of all the available LHC Higgs data so far, and derive $2\sigma$ ($1\sigma$) 
upper limits of 0.25 (0.15) on the invisible Higgs decay branching ratio in this scenario. These in turn put upper limits of order 0.1 on the Dirac Yukawa coupling in this model. We further show that the model parameter space allowed by the invisible Higgs decay branching ratio limits 
is fully accessible in the near future DM direct detection experiments such as LUX and XENON1T, and can be ruled out {\it completely} in case of a null result from these experiments. Finally, we 
have explored the prospects of the invisible Higgs decay signature at the $\sqrt s=14$ TeV LHC for a chosen set of benchmark points. We find that a signal significance of $3\sigma$ can be achieved 
in the VBF channel with an integrated luminosity as low as $200~{\rm fb}^{-1}$, whereas in the $Zh$ channel, it requires a luminosity of at least $600~{\rm fb}^{-1}$ for our chosen benchmark points.
\section*{Acknowledgment}
We are extremely thankful to the anonymous referee for carefully reading the whole manuscript, 
making many valuable suggestions, and checking some of the numerical results.
We thank  Sanjoy Biswas, Pushan Majumdar, Dipan Sengupta, and Florian Staub for 
useful correspondence, suggestions and comments at different stages of the 
work. SB and SM wish to thank Arindam Chatterjee and Satyanarayan Mukhopadhyay for some useful suggestions and comments. 
The work of PSBD is supported by the 
Lancaster-Manchester-Sheffield Consortium for Fundamental Physics under STFC 
grant ST/J000418/1.  
SM acknowledges the hospitality of RECAPP, (HRI), Allahabad that led to this fruitful collaboration, 
and also wishes to thank the Department of Science and Technology, Government of 
India for a Senior Research Fellowship. S.B. and B.M. thank the Indian Association for the Cultivation 
of Science, Kolkata, for hospitality while this project was in progress.
The work of SB and BM was partially supported by funding available from the Department of Atomic Energy, Government
of India for the Regional Centre for Accelerator-based Particle Physics, Harish-Chandra Research Institute. 
Computational work for this study was partially carried out at the cluster
computing facility in the Harish-Chandra Research Institute (http://cluster.hri.res.in).
 
\section*{Appendix: Higgs Data Sets}
In Table~\ref{tab:tab1}  we list the latest Higgs data sets available from the combined $\sqrt s = 7$ and 8 TeV 
LHC run in five visible Higgs decay channels: $\gamma\gamma,~ZZ^*\to 4\ell,~WW^*\to 2\ell 2\nu,~ b\bar{b}$ and $\tau\bar{\tau}$.  For each channel, we show the experimental values of the signal strengths $\hat{\mu}_i$ together with its $1\sigma$ uncertainty, as reported by the ATLAS and CMS collaborations~\cite{Aad:2013wqa, cmsg, cmsz, cmsw, atlasb, cmsb, atlastau, cmstau}.  
\begin{table}[htb]
\centering
\begin{tabular}{c|c|c}
\hline\hline
Channel & $\hat{\mu}$ & Experiment \\
\hline \hline
  $h \to \gamma \gamma$ & $1.55^{+0.33}_{-0.28}$ & ATLAS~\cite{Aad:2013wqa} \\
  & $0.78^{+0.28}_{-0.26}$ & CMS~\cite{cmsg} \\
\hline
  $h \rightarrow Z Z^{*} \to 4l$ & $1.43_{-0.35}^{+0.40}$ & ATLAS~\cite{Aad:2013wqa} \\
 & $0.9_{-0.20}^{+0.30}$ & CMS~\cite{cmsz} \\
\hline
  $h \rightarrow W W^{*} \to 2l 2\nu$ & $0.99_{-0.28}^{+0.31}$ & ATLAS~\cite{Aad:2013wqa} \\
& $0.80_{-0.20}^{+0.20}$ & CMS~\cite{cmsw} \\
\hline
  $h \rightarrow b \bar{b}$ & $0.20_{-0.60}^{+0.70}$ & ATLAS (VH)~\cite{atlasb} \\
& $1.00_{-0.50}^{+0.50}$ & CMS (VH)~\cite{cmsb}\\
\hline
  $h \rightarrow \tau \bar{\tau}$ & $0.7_{-0.6}^{+0.7}$ & ATLAS~\cite{atlastau} \\
& $1.10_{-0.4}^{+0.4}$ & CMS~\cite{cmstau} \\
\hline
\hline
 \end{tabular}
\caption{Data set used in our analysis, with the values of $\hat{\mu_i}$
  in various channels and their $1\sigma$ uncertainties as reported by the
  ATLAS and CMS collaborations.}
\label{tab:tab1}
\end{table}

\end{document}